\documentclass{ar-1col-S2O}
\usepackage[numbers]{natbib}
\usepackage{url}

\usepackage{amsmath}
\usepackage{amssymb}
\usepackage{amsthm}
\usepackage{amscd}
\usepackage{amsfonts}
\usepackage{graphicx}%
\usepackage{fancyhdr}
\usepackage{caption}
\usepackage{subcaption}
\usepackage{wrapfig}
\usepackage{helvet}
\usepackage{stmaryrd}
\usepackage{bm}
\usepackage[table]{xcolor}
\usepackage{booktabs}
\usepackage{setspace}
\usepackage{wrapfig}
\usepackage{hyperref}

\usepackage{soul}

\jname{Xxxx. Xxx. Xxx. Xxx.}
\jvol{AA}
\jyear{YYYY}
\doi{10.1146/((please add article doi))}

\newcommand{\Up}{\mathbf{U}}
\newcommand{\Dn}{\mathbf{D}}

\begin{document}

\markboth{Mungan \& Sastry}{Self-organization in driven disordered systems}

\title{Self-organization, Memory and Learning: From Driven Disordered Systems to Living Matter}

\author{Muhittin Mungan$^{1,2}$, Eric Cl{\' e}ment$^{3,4}$, Damien Vandembroucq$^4$ and Srikanth Sastry$^5$
\affil{$^1$Institute for Biological Physics, University of Cologne, Cologne, Germany; email: mungan@thp.uni-koeln.de}
\affil{$^2$Institute of Theoretical Physics II, Heinrich-Heine University, Dusseldorf, Germany; email: muhittin.mungan@hhu.de}
\affil{$^3$Institut Universitaire de France (IUF), Paris, France; email: eric.clement@upmc.fr}
\affil{$^4$PMMH, CNRS, ESPCI Paris, Universit\'e PSL, Sorbonne Universit\'e, Universit\'e Paris Cit\'e, France; email: damienvdb@pmmh.espci.fr}
\affil{$^5$Jawaharlal Nehru Centre for Advanced Scientific Research, Jakkur Campus, Bengaluru, India, 560064; email: sastry@jncasr.ac.in}
}

\begin{abstract}
%
Disordered systems subject to a fluctuating environment can self-organize into a complex history-dependent response, retaining a memory of the driving. In sheared amorphous solids, self-organization is established by the emergence of a persistent system of mechanical instabilities that can repeatedly be triggered by the driving, leading to a state of high mechanical reversibility. As a result of self-organization, the response of the system becomes correlated with the dynamics of its environment, which can be viewed as   a sensing mechanism of the system's environment. Such phenomena emerge across a wide variety of soft matter systems, suggesting that they are generic and hence  may depend very little on the underlying specifics. We review self-organization in driven amorphous solids, concluding with a discussion of what self-organization in driven disordered systems can teach us about how simple organisms sense and adapt to their changing environments.
\end{abstract}

\maketitle

\tableofcontents

\section{Introduction}

The theme of this review addresses the following question: {\em What can self-organization, memory formation and learning in a driven disordered system teach us about how simple organisms lacking a brain sense and adapt to their changing environment?} Our aim is to present a theoretical understanding of how self-organization and learning emerge in a sheared amorphous solid. We then describe how such insights may guide us in understanding how simple organisms, like bacteria, sense and adapt to changing environments. In this first section we present the rich and complex response of driven disordered systems, and conclude with a detailed outline of this review's contents and its key points. %

Understanding the response of a disordered solid to an externally imposed forcing,
such as stress or strain, is important in order to characterize the transitions between rigid and flowing states in a
wide variety of soft matter systems. Examples for such
behavior include the jamming transition in granular materials~\cite{behringer2018physics}, the yielding transition in amorphous solids~\cite{bonn2017yield,nicolas2018deformation}, the hysteretic response of crumpled thin elastic sheets \cite{shohat2022} and the depinning transition of a pinned elastic interface, such as flux-lines in type II superconductors~\cite{Blatter}. The interplay between deformation energy cost and gain, as the disordered solid adapts to the imposed forcing, gives rise to rich dynamics on a complex energy landscape, exhibiting dependence on deformation history and self-organization.

\begin{figure}[t!]
\begin{center}
\includegraphics[width=\columnwidth]{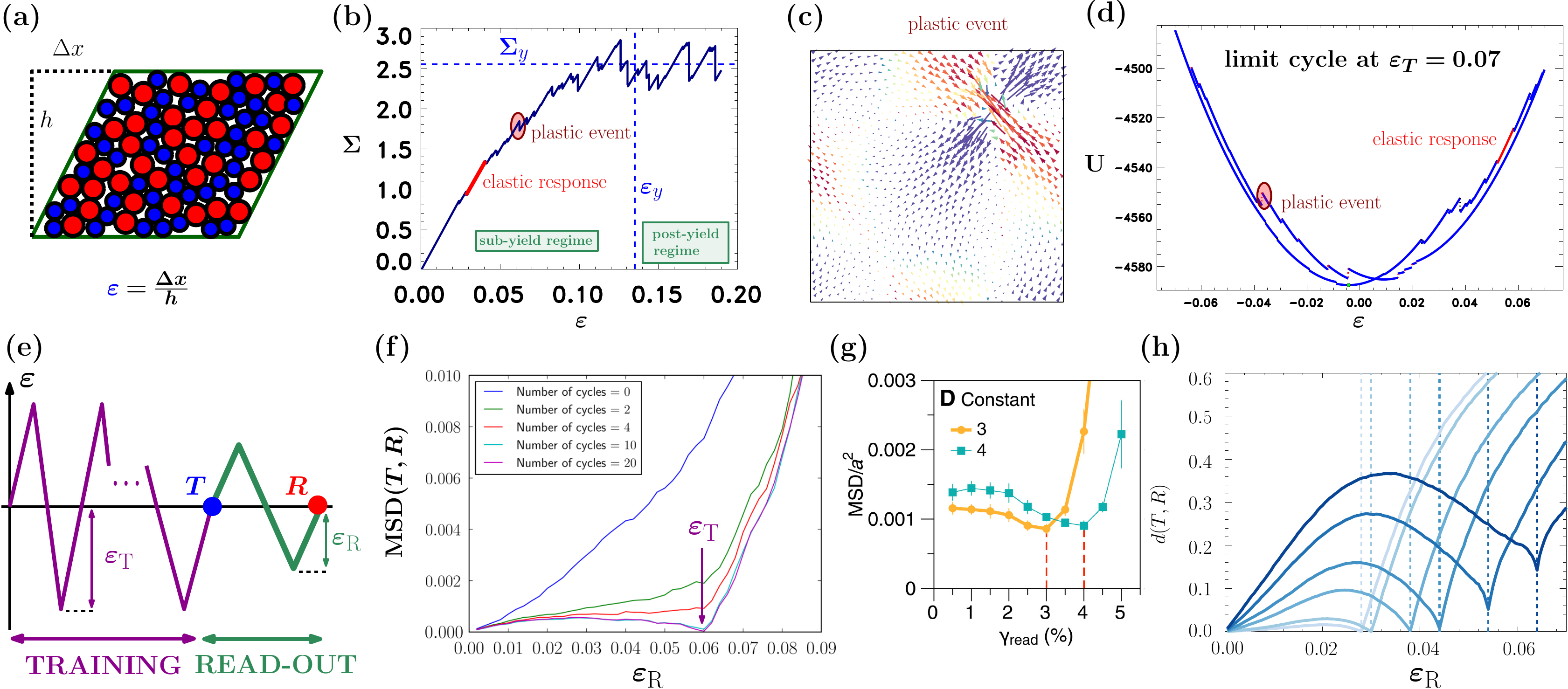}
\caption{{\bf The phenomenology of yielding and memory formation in an amorphous solid under athermal quasistatic (AQS) shear.} {\bf (a)} Binary mixture of particles in a 2d box undergoing shear deformation at strain $\varepsilon$. {\bf (b)} Stress-strain response of the system in (a) under uniform increasing strain. The  build-up of stress $\Sigma$ is punctuated by stress-drops due to plastic events,  continuing until the system yields and $\Sigma$ fluctuates around  $\Sigma_y$, the yield stress. {\bf (c)} Plastic events are caused by spatially localized rearrangements of particles (shaded red area), called shear transformations or soft-spots. {\bf (d)} Steady-state behavior of the configurational potential energy due to application of cyclic shear at amplitude $\varepsilon = 0.07$, resulting in a microscopically periodic response with repeatedly and reversibly triggered plastic events.
{\bf (e)} Illustration of a training protocol of  $\mathcal{N}$ shearing cycles applied at amplitude $\varepsilon_T$, leading to the trained state $T$. This is followed by a read-out where a single shear cycle at amplitude $\varepsilon_R$ is applied to copies of $T$, and a distance $d(T,R)$ between the particle configurations at $T$ and  $R$ is computed. {\bf (f)} Evolution of $d(T,R)$ as a function of $\varepsilon_R$ for various durations $\mathcal{N}$ of training. The kink around $\varepsilon_T \approx \varepsilon_R$ develops already after a few training cycles. {\bf (g)} Experimentally obtained read-out curve obtained upon training a system of polysterene particles confined to an oil-water interface.
{\bf (h)} Read-out response from training a mesoscale model of an amorphous solid at various amplitudes $\varepsilon_T$, as color coded in darkening shades of blue and 
indicated by dashed verticals.    
Panels (b) -- (d) adapted or plotted from data obtained in \cite{mungan2019networks}, (f) adapted from \cite{fiocco2014encoding}, (g) adapted from \cite{keim2022ringdown} (CC BY 4.0), (h) adapted from \cite{kumar2024self}.}       

\label{fig:amorphous-intro}
\end{center}
\end{figure}

Fig.~\ref{fig:amorphous-intro}(a) shows the athermal response of an amorphous solid to deformations by an externally applied shear strain. When subject to monotonously increasing shear $\varepsilon$, the amorphous solid responds by a stress $\Sigma$ opposing the deformation, as seen in Fig.~\ref{fig:amorphous-intro}(b). The response under increasing strain $\varepsilon$ is characterized by stress-strain segments in which $\Sigma$ builds up continuously. These segments correspond to purely elastic changes of particle configurations and are punctuated by stress drops due to development of mechanical instabilities. The latter are {\em plastic events} and lead to abrupt spatially localized rearrangements, called {\em shear transformation zones} or {\em soft-spots} \cite{argon1979plastic,falk1998dynamics,manning2011softspot,Richard-PRM20}\footnote{Note that in the following our use of the term ``soft-spot", historically associated with the localized eigenmode of a soft vibrational mode, {\it i.e.} characterized by a low eigenvalue that approaches zero as a result of shear, will encompass the notions of plastic events, shear transformations, and local rearrangements. The term ``hot spot" has also been used, as we describe below, to discuss plastic rearrangements. We prefer to use the term ``soft spot'' in this context as well. If a distinction were to be made, soft spots may be considered as the immediate precursors of a plastic event and hot spots and local rearrangements as their actual (numerical or experimental) realization (See ref.~\cite{Richard-PRM20} for a thorough discussion about plastic events and their predictors).}.  Fig.~\ref{fig:amorphous-intro}(c) illustrates the particle displacements occurring during a typical plastic event. The region shaded in red shows a  shear transformation. The accompanying particle displacements (indicated by vectors) exhibit a characteristic quadrupolar pattern and give rise to long-range elastic interactions via redistribution of local stresses. Increasing $\varepsilon$ further, $\Sigma$ settles into a steady-state, fluctuating around a value $\Sigma_y$,  called the {\em yield stress}, which marks the onset of {\em yielding}.

Experiments on colloidal suspensions and simulations of amorphous solids show that when subject to oscillatory shear strain at 
amplitude $\varepsilon_T$, these systems can settle into a periodic response \cite{keim2014mechanical,Royer49,regev2013onset,fiocco2013oscillatory, mungan2019networks, keim2022ringdown}.
Moreover under athermal conditions, numerical simulations reveal that this response is {\em microscopically} periodic, meaning that the particles return to their initial positions at the end of each period. Fig.~\ref{fig:amorphous-intro}(d) shows the typical behavior of the potential energy $U$ during such a limit cycle.
Again, the response is composed of purely elastic deformations, described by the parabolic segments of the configurational potential energy, which are interrupted by energy drops that accompany plastic events. Experiments and simulations further show that as the amplitude of oscillatory shear $\varepsilon_{T}$ is increased,  the number of driving cycles needed to reach a cyclic response increases, the cyclic response gets increasingly more complex, consisting of a larger number of plastic events, and its  period starts to span multiple driving cycles. Beyond a critical amplitude $\varepsilon_{\rm c}$, cyclic response is not attainable anymore,  marking the onset of the {\em irreversibility transition} \cite{regev2013onset}. 

The process of subjecting the amorphous solid to cyclic shear at strain amplitude $\varepsilon_T$ for a number $\mathcal{N}$ of cycles,  anneals it mechanically. Denoting by $T$ the state of the glass ``trained'' by such  annealing, it turns out that $T$ encodes a {\em memory} of  $\varepsilon_T$ that can subsequently be read-out. This is illustrated in Fig.~\ref{fig:amorphous-intro}(e): after training, copies of $T$ are subjected to single shear cycles with different read-out amplitudes $\varepsilon_R$. Comparing the trained glass configuration $T$ with the configurations $R$ reached after the application of a read-out cycle  at various strains $\varepsilon_R$, shows that when the training and read-out amplitudes are close to each other, i.e. $\varepsilon_T \approx \varepsilon_R$, so are the corresponding configurations \cite{keim2022ringdown}. This is shown in Fig.~\ref{fig:amorphous-intro}(f), where the particles mean squared displacement (MSD) between  $T$ and $R$ is plotted against read-out amplitudes $\varepsilon_R$ for various numbers $\mathcal{N}$ of training cycles. Already after $10$ driving cycles, the MSD curve  develops a dip where training and read-out amplitudes coincide. The encoding and read-out of such  memories has been observed experimentally for various soft-matter systems, including sheared colloids and  crumpled elastic sheets under compression \cite{keim2019memory, keim2018return, shohat2022, keim2022ringdown, paulsen2024mechanical}. 
Fig.~\ref{fig:amorphous-intro}(g) shows an experimental read-out result after training a system of polysterene spheres confined to an oil-water interface\footnote{Experimentally, copies of  $T$ may not be available. Read-out is performed by applying cycles with successively increasing amplitude} \cite{paulsen2024mechanical}. \cite{keim2022ringdown}, while  Fig.~\ref{fig:amorphous-intro}(h) shows 
similar memory retrieval from a mesoscale model of a sheared amorphous solid \cite{kumar2024self}.

We shall argue that the mechanism of memory formation and self-organization in soft matter systems may be relevant for understanding adaptation of simple organisms lacking a nervous system. Sensing changes by processing environmental inputs and reacting to these is essential for the survival of organisms. Such adaptation must involve some form of memory, learning and information processing \cite{libchaber2020walking,stern2023learning,floyd2024limits,tagkopoulos2008predictive,Murugan_2021, gershman2021learning,vermeersch2022microbes}. For example, the cytoskeleton of eukaryotic cells can dynamically adjust its physical properties according to mechanical deformations exerted by its environment \cite{gralka2015biomechanics}. 

The  outline of this review is as follows. A natural way to approach memory and self-organization in disordered solids is to focus on two key concepts: hysteresis and multistability. In Section \ref{sec:so-amorphous} we introduce the tools and concepts recently developed in this context. The notions of $t$-graphs, which describe transitions between metastable states, and return-point memory provide a solid framework within which the memory behavior in disordered solids can be described. The emergence of mechanical reversibility under annealing is due to a system of repeatably triggerable instabilities, the soft-spots, which are spatially-localized in space and exhibit hysteresis. We discuss the role of frustration in soft-spot interactions and then make an excursion to the Preisach model, which can be thought of as an "ideal gas" of non-interacting soft-spots. The Preisach model illustrates how driving histories are encoded as memories via its elementary units of hysteresis, the hysterons. 

The mapping of mechanical deformation histories onto paths on a transition graph enables us to directly associate mechanical reversibility with the topology of the graph: limit cycles and any other repeatable behavior can exist only within its strongly connected components, the regions of mutual reachability of the graph. 
In Section \ref{sec:self-organization}, we show that the formation of memory is robust and need not depend on the nature of driving. Random driving on large length scales, mimicking environmental fluctuations, or correlated on short time and length scales, resembling active matter, lead to similar features of memory formation.  

We next look in Section \ref{sec:to_live_or_not} more closely at the process of self-organization.  We show that self-organization under mechanical annealing, be it random or deterministic, leads to the emergence of a persistent soft-spot system, that very much like the idealized Preisach system can encode memories of deformation histories. We can therefore view self-organization in a fluctuating environment as a pathway by means of which an internal representation of the  environment is established.

Armed with these insights, we address in Section \ref{sec:bio} the question raised at the beginning, namely  how memory and self-organization in driven disordered systems can guide us in understanding how simple organisms adapt to changing environments. This section is necessarily incomplete and serves to motivate future research. We first review the biological evidence of history-dependent and anticipatory behavior in simple organisms, discussing the role of biological switches, and multistability. We make the connection between non-living and living systems via the notion of dimensional reduction and soft-modes, which are key ideas underlying the theoretical descriptions in both realms. The $t$-graph description in driven disordered systems, leading to a focus on soft-spots, is an example of dimensional reduction and the emergence of soft-modes. 
This is followed by an exploration of the parallels between mechanical annealing and adaptive evolution. We conclude with a brief outlook.  

The topics of three recent Annual Reviews complement the ideas and material presented here: exploring memory formation and its applications in metamaterials \cite{paulsen2024mechanical}, learning in driven disordered systems \cite{stern2023learning} and the role of soft-modes in biology \cite{russo2025softmodes}.

\section{Self-organization in driven disordered systems: The sheared amorphous solid}
\label{sec:so-amorphous}

The response of amorphous solids to mechanical deformations described above is not  fully understood even under athermal and quasistatic (AQS) conditions.  This is essentially due to a lack of crystalline order in these amorphous materials, 
and consequently lack of well identified structural defects
giving rise to plastic events.   
While this problem attracted the interest of material scientists since the late 1970s \cite{argon1979plastic}, it is only since the 1990s that it also received the attention of theoretical physicists \cite{sollich1997rheology,falk1998dynamics,hebreaudlequeux1998}. By now, there are theoretical models  capturing  the irreversibility transition under cyclic shearing, such as the approach of  \cite{parley2022mean} based on an adaptation of the 
H{\' e}braud-Lequeux model \cite{sollich1997rheology} (See also \cite{chenliu2020, kumar2022, suda2025yieldingmemorydrivenmeanfield}), as well as an analytically tractable mesostate model \cite{sri2021mesoland, mungan2021metastability}. These latter models capture qualitatively the ``phase diagram'',  describing the response and its dependence on sample aging and oscillatory shear amplitude. However, owing to their mean-field nature, these models do not incorporate any spatial  structure and fail to account for microscopic plastic reversibility.

\subsection{The transition graph description of the AQS response}
\label{subsec:t-graph}

Our starting point is the observation that the mechanical response of an amorphous solid to external deformation under AQS conditions consists of series of elastic segments interspersed by instabilities. Such a sequence of transitions between metastable states  has a natural representation in terms of a transition graph ($t$-graph)~\cite{munganterzi2018, munganwitten2018, mungan2019networks}.  The dynamical features of the AQS response are thereby encoded faithfully in the $t$-graph {\em topology}, providing new insights into these phenomena \cite{munganterzi2018, mungan2019networks,terzi2020state,Regev2021, mungan2022comm, kumar2022, kumar2024self, mungan2024self}. 

\begin{figure}[t!]
\begin{center}
\includegraphics[width=\textwidth]{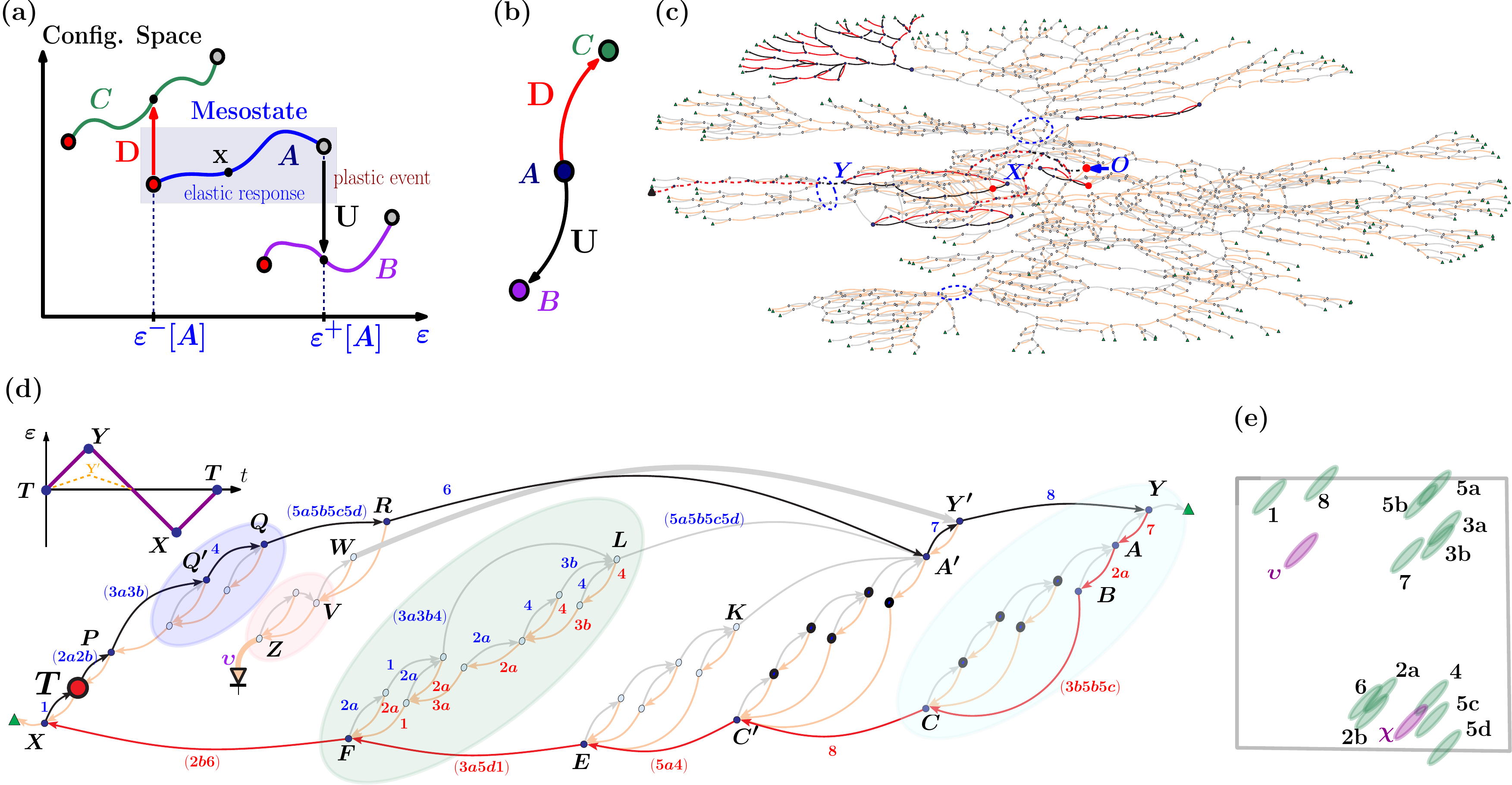}
\caption{{\bf Transition graph-description of the AQS response of an amorphous solid subject to shear.}
{\bf (a)} Viewed in the configuration space of particles, the AQS response to external shear $\varepsilon$ leads at first to continuous deformations of particle configurations $\bm{x}$, tracing out a hypercurve segment in configuration space, called an  {\em elastic branch}. Its termination points mark the onset of mechanical instabilities which lead to other elastic branches, e.g.  transitions from elastic branch $A$ to $B$ and $C$. Here and in the following we will denote elastic branches by capital italic letters while we will reserve the bold face letters $\Up$ and $\Dn$, in black and red respectively, to indicate the transition from the upper and lower stability limit of an elastic branch.  {\bf (b)} The transitions can be represented as a directed graph whose vertices are the elastic branches. {\bf (c)} Transition graphs ($t$-graphs) can be sampled from standard molecular dynamics simulations and reveal complex topological features. 
Each deformation protocol corresponds to a path on the graph. 
The evolution towards periodic response under cyclic shear applied to the glass $O$ at various amplitudes are shown by the dashed pathways. {\bf (d)} The limit cycle reached from $O$ with training amplitude $\varepsilon_T = 0.05$.  Top left: the training by oscillatory shear leads to the periodic response $T \to Y \to X \to T$. Center: Detailed view of $t$-graph associated with the $0.05$ limit-cycle marked in (c). Transitions under increasing and decreasing strain are shown as black/gray and red/orange arrows, respectively. 
The periodic response corresponds to the graph cycle traced out by starting from the vertex $X$, following the black arrows until $Y$, and subsequently following the red arrows back to $X$. The graph-cycles are labeled by their extreme points, e.g. $(X,Y)$, and exhibit hierarchical nesting, e.g. $(F,L)$ within $(F,Y')$ within $(X,Y')$ within $(X,Y)$. Nesting is characteristic of loop return-point-memory \cite{munganterzi2018}. {\bf (e)} The transitions forming the main hysteresis cycle $(X,Y)$ are due to $13$ soft spots whose sample locations are indicated by green ellipses. The labels on some transitions shown in  (d) indicate the soft-spots involved. Panels (c) -- (e) adapted from \cite{mungan2019networks}.
} 
\label{fig:tgraph-intro}
\end{center}
\end{figure}

Consider the AQS dynamics of a $2d$ atomistic amorphous solid that is subjected to shear strain $\varepsilon$, as depicted in Fig.~\ref{fig:amorphous-intro}(a). After its initial preparation, and before the application of shear, the particle configuration resides in a local minimum ${\bf x}$ of its high-dimensional potential energy surface $U({\bf x}, \varepsilon)$, as schematically illustrated in Fig.~\ref{fig:tgraph-intro}(a). As we increase the strain in a slow and adiabatic manner, the energy landscape deforms and the position of the local energy minimum moves along a hypercurve in configuration space, as indicated by the blue curve.
The amorphous solid adapts to such strain increments by purely elastic deformation. This elastic response continues until a strain at which the particle configuration ceases to be a local energy minimum and thus becomes unstable. Increasing the strain slightly further, the system relaxes into a new local energy minimum and this constitutes a plastic event. Thus, given a locally stable particle configuration ${\bf x}$, there exists a range of strains, applied in the positive and negative directions, over which the amorphous solid adapts in a reversibly elastic manner and which is punctuated in either direction by plastic events\footnote{This is the implicit function theorem in disguise: mechanical equilibrium requires $\nabla_{\bf x}U({\bf x}, \varepsilon) = 0$, and  implicitly defines curve segments ${\bf x}(\varepsilon)$ in configuration space.
}. We refer to these curve segments as {\em mesostates} \cite{mungan2019networks}. They correspond to the minimized structures of \cite{heuer2018mesostate} and are also called {\em elastic branches}. 

With each mesostate $A$ we associate the range of strain values $(\varepsilon^-[A], \varepsilon^+[A])$, over which the locally stable configurations transform elastically into each other. Plastic events occur at $\varepsilon^\pm[A]$ whereby the system reaches  new locally stable configurations which must necessarily belong to some other mesostates, as indicated by the black and red arrows in Fig.~\ref{fig:tgraph-intro}(a). Hence mesostate transitions are triggered at either end of the stability interval $(\varepsilon^-[A], \varepsilon^+[A])$. We distinguish these by referring to transitions under strain increase and decrease as $\Up$-, respectively, $\Dn$-transitions. The transitions from $A$ to $B$ or $C$ depicted in Fig.~\ref{fig:tgraph-intro}(a) can be represented by a directed graph, as shown in Fig.~\ref{fig:tgraph-intro}(b) and  lead to a description of mesostate transitions in terms of a {\em transition graph}, the {\em $t$-graph} \cite{munganterzi2018, munganwitten2018, mungan2019networks}. Here each vertex represents a mesostate $A$, from which there are two outgoing directed transitions, 
$\Up$ and $\Dn$, describing the outcome of the plastic events taking place at $\varepsilon^\pm[A]$.   

Fig.~\ref{fig:tgraph-intro}(c) shows an excerpt of a $t$-graph that was extracted from molecular dynamics of a $2d$ amorphous solid of $1024$  particles \cite{mungan2019networks}. The undeformed glass $O$ is the red vertex, while the $t$-graph shows all $1416$ mesostates reachable from $O$ by arbitrary sequences of  $\Up$- and $\Dn$-transitions of maximal length $g = 25$. Here and in the following we will adopt the convention to represent $\Up$- and $\Dn$-transitions by arrows colored  in gray resp. orange, while colors black and red will be used to highlight transitions of interest.
Note that the $t$-graph is non-homogeneous, there are tree-like regions, highly connected regions and bottlenecks (blue ovals), i.e. certain parts of the graph can be reached only  through specific vertices.

The conventional way of extracting dynamical features from numerical simulations typically focuses on macroscopic quantities extracted from time-series, such as the responses in Fig.~\ref{fig:amorphous-intro}.
The  $t$-graph description of such systems provides a complementary way of obtaining relevant information  \cite{mungan2019networks, Regev2021, kumar2022}. It gives a bird's eye view of the possible mechanical responses of an initially prepared glass $O$, since the result of applying to $O$ {\em any} deformation protocol, can be traced out by following the corresponding $\Up$- or $\Dn$-transitions on the graph.
In this way, features of the complex AQS response are encoded into the $t$-graph topology: the yielding and the irreversibility transitions are all on this graph -- somewhere.   

\subsection{Return-point memory} 
\label{subsec:rpm}
In Fig.~\ref{fig:tgraph-intro}(c) the sequences of black and red dashed transitions starting from $O$ indicate deformation pathways under cyclic shear at various strain amplitudes $\varepsilon_{T}$ \cite{mungan2019networks}. The graph-cycle marked with the endpoints $X$ and $Y$ is the limit cycle reached from $O$ when $\varepsilon_{T} = 0.05$ (the onset of the irreversibility transition for this system is  $\varepsilon_c = 0.13$).  Fig.~\ref{fig:tgraph-intro}(d) is a detailed view. 
Under the   
strain $0 \to  \varepsilon_{T} \to - \varepsilon_{T} \to 0$, the 
system traces out the graph-cycle $T \to Y \to X \to T$, following the black (red) arrows under strain increase (decrease). The mesostates 
$X$ and $Y$ reached at the extreme points $\pm \varepsilon_T$ of the driving are the lower and upper endpoints of the graph-cycle $(X,Y)$ \cite{munganterzi2018}. 

In fact, the $t$-graph excerpt shown is complete, since it contains all possible $\Up$- and $\Dn$-transition out of each of the mesostates associated with the graph-cycle $(X,Y)$. The $\Up$-transition out of $Y$, the $\Dn$-transition out of $X$, as well as a ``rabbit hole'' $\Dn$-transition \cite{mungan2019networks} out of $Z$, indicated by the diode sign,  lead to states ``outside'' of the cycle $(X,Y)$, a statement we will make more precise shortly.  We call the subgraph shown in Fig.~\ref{fig:tgraph-intro}(d) the {\em loop-graph} associated with $(X,Y)$ \cite{munganterzi2018}. Remarkably, the loop-graph $(X,Y)$ exhibits hierarchical nesting of graph-cycles within cycles. One such  example is the graph-cycle $(F,L)$ outlined by the green ellipse in Fig.~\ref{fig:tgraph-intro}(d), which is a sub-cycle of $(F,Y')$, which in turn is a sub-cycle of $(X,Y')$, {\em etc.} 

The topological nesting of hysteresis cycles is a characteristic feature of return-point-memory (RPM). Its $t$-graph-topological manifestation has been called {\em loop return-point-memory} ($\ell$RPM) and was defined and analyzed in \cite{munganterzi2018}. Roughly,  $\ell$RPM prescribes that {\em mesostates belonging to a loop-graph can leave it only through its endpoints}. As an illustration, consider the loop-graph $(F,L)$. Any mesostate inside this loop, i.e. inside the shaded green area, must go first through $F$ or $L$ before it can reach any other mesostate outside of the loop\footnote{The $\Dn$-transition out of $Z$ violates $\ell$RPM, which requires that following the $\Dn$-arrows from $Z$,  $X$ must eventualy be reached \cite{mungan2019networks}. The diode-sign indicates the one-way nature of this transition.}.
The placement of loop-graph vertices  highlights the $\ell$RPM structure.

RPM is a feature associated with ferromagnets and more generally systems with interactions where the instability of one part has a destabilizing effect on other parts \cite{middleton1992asymptotic, sethna1993hysteresis, munganterzi2018}.  Due to the quadrupolar nature of the elastic interactions in amorphous solids \cite{eshelby1957determination,argon1979plastic}, there is also frustration: instabilities in one part of the system may not only destabilize but also stabilize other parts. One would therefore not expect sheared amorphous solids to exhibit RPM, and the hierarchical nesting of cycles observed comes as a surprise. In fact, loop-graphs extracted from simulations exhibit near perfect $\ell$RPM, even at strain amplitudes where interactions between soft-spots cannot be ignored \cite{mungan2019networks}. These findings are consistent with  experiments  
that probe memory formation and RPM in non-Brownian suspensions \cite{corte2008random}, sheared colloids \cite{keim2014mechanical,keim2018return}, crumpled elastic sheets \cite{shohat2022}, and simulations of sheared amorphous solids \cite{regev2013onset,fiocco2014encoding}.

The nesting of graph-cycles also explains the memory formation upon read-out, as shown in Figs.~\ref{fig:amorphous-intro}(f) and (g). Starting at $T$ and applying cyclic shear at read-out amplitudes $\varepsilon_R$, will confine the response to a sub-graph of the loop-graph $(X,Y)$ from which it is possible to return to the trained hysteresis cycle, as long as the applied strains are such that we avoid exiting the loop graph via $X,Y$ or $Z$.

\subsection{Emergence of mechanical reversibility} 
\label{subsec:mechrev}

It is apparent that the applied oscillatory strain not only has caused the system to evolve towards a cyclic response, but at the same time the response has {\em self-organized} itself into a hierarchy of nested sub-cycles.
These features generalize to the graph-topological notion of {\em mutual reachability}: a pair of mesostates $A$ and $B$ is mutually reachable, if there exist  deformation pathways from $A$ to $B$ as well as from $B$ to $A$. Mutual reachability is an equivalence relation and partitions the vertex set of the $t$-graph into equivalence classes, the {\em strongly connected components} or SCCs \cite{barrat2008dynamical}. 
\begin{marginnote}[]
\entry{Equivalence Relation}{A relation $\sim$ between pairs  $(a,b)$ satisfying: (i) $a \sim a$, (ii) $a \sim b$ is equivalent to $b \sim a$ and (iii) if $a \sim b$ and $b \sim c$ then $a \sim c$.  
}
\end{marginnote}
Observe that all mesostates shown in 
Fig.~\ref{fig:tgraph-intro}(d) are mutually reachable and hence part of a single SCC. In fact, any mechanically reversible response must be confined to a single SCC, as transitions out of an SCC are by definition irreversible. The partition into SCCs allows us to classify the plastic events accompanying the transitions of the $t$-graph as being {\em plastically reversible} (intra-SCC transitions) or {\em plastically irreversible} (inter-SCC transitions)  \cite{Regev2021}.  

The transitions shown in Fig.~\ref{fig:tgraph-intro}(d) are plastic events that are due to 13 soft-spots whose locations in the sample have been marked by the green ellipses in Fig.~\ref{fig:tgraph-intro}(e). These soft-spots turn out to be two-state systems that hysteretically switch between states as the applied strain is increased or decreased. In fact, they emulate the hysterons of the Preisach model which we discuss in Section \ref{subsec:preisach}. 
The blue and red labels next to some transitions in Fig.~\ref{fig:tgraph-intro}(d) indicate soft-spots changing states. Note how a subset of the soft-spots is active in sub-cycles of the $t$-graph, such as those in the loop-graph $(F,L)$ highlighted in green. 
\begin{marginnote}[]
\entry{Preisach Hysteron}{A two-state system with hysteresis: the actual state occupied depends on the driving history. 
}
\end{marginnote}

Soft-spots interact with each other via the long-range elastic-deformations triggered by their state changes which cause redistribution of local stresses, cf. Fig.~\ref{fig:amorphous-intro}(c). As a result, soft-spots can alter the switching behavior of other soft-spots, they can give rise to new ones or disable some (by making the corresponding region respond elastically). The mechanical annealing by cyclic shear has thus led to self-organization via the {\em selection} of an interacting soft-spot system which is also {\em persistent} \cite{mungan2022comm}: soft-spots are not only active when traversing the limit-cycle, but subsets can selectively change states in sub-cycles. This persistency in turn gives rise to mechanically reversible behavior, whose extent is captured by the SCCs. SCCs ultimately facilitate memory formation by making possible mechanical driving protocols that can bring back the system to its trained state. SCC sizes and their accessibility via mechanical deformations characterizes the extent to which a system responds in a mechanically reversible manner \cite{Regev2021, kumar2022}.

\subsection{The role of frustration in mechanical annealing} 
\label{subsec:frustration}

We turn next to the role of frustration which is part of the soft-spot interactions. To motivate this, consider the ``rabbit-hole'' transition out of $Z$ in Fig.~\ref{fig:tgraph-intro}(d). This is a case where the interactions between soft-spots have led to the creation of a new soft-spot, labeled $\upsilon$ in panels (d) and (e). The creation of this soft spot has altered the particle configuration in an irreversible manner, leading to a premature exit from the loop-graph $(X,Y)$. That is, without passing through its endpoints $X$ or $Y$, which would have required the application of strains with magnitudes larger than the training strain $\varepsilon_T$. The rabbit-hole transitions that end up by-passing the endpoints of a loop are a consequence of frustration, which is a characteristic of soft-spot interactions. If the return point memory property was obeyed such ``rabbit-hole'' transitions could not occur \cite{sethna1993hysteresis, munganterzi2018}.  

The classical route to return point memory is through a   partial ordering of mesostates that is preserved under the application of driving. It is called the no-passing (NP) property\footnote{Let $A_t$ and $B_t$ be the evolutions of mesostates $A_0$ and $B_0$ under driving forces $f_A(t)$ and $f_B(t)$. The dynamics obeys the NP property, if there exists a partial order $\preceq$ among mesostates, such that when $A_0 \preceq B_0$, and $f_A(t) \le f_B(t)$ over a time interval $0 \le t \le T$, then $A_t \preceq B_t$.} and was formulated first in the context of elastic manifolds driven through random media \cite{middleton1992asymptotic}. 
\begin{marginnote}[]
\entry{Partial Order}{A set of objects is partially ordered, if there exists a relation $\precsim$ such that (i) $A \precsim A$, (ii) if $A \precsim B$ and $B \precsim A$ then $A = B$ and (iii) $A \precsim B$ and $B \precsim C$ implies $A \precsim C$.  
}
\end{marginnote}
NP implies RPM, as was proven under general conditions in \cite{sethna1993hysteresis} and reviewed within the context of mesostates and $t$-graphs in \cite{munganterzi2018}. The mechanism underlying NP is related to how the triggering of an instability affects the switching thresholds of the other instabilities. In the context of magnets and scalar spins, ferromagnetic interactions implies NP, thereby leading to RPM \cite{sethna1993hysteresis}. The sidebar below explores consequences of NP on the $t$-graph topology.
\begin{marginnote}[]
\entry{Ferromagnetic Partial Order}{Consider a magnet with  spins  either up or down. Spin configurations $A$ and $B$ are partially ordered as $A \precsim B$, if every up-spin in $A$ is also an up-spin in $B$.
}
\end{marginnote}

Antiferromagnetic interactions lead to frustration and therefore tend to break NP, but this is not necessarily always the case \cite{Deutschetal2004}. However, when frustration is present and NP does not hold, then (i) SCCs do not have to be loop-graphs anymore, (ii) they can be exited prematurely via rabbit holes, and (iii) it may take more than one driving cycle to land in a trapping SCC, if at all. Thus  interactions that give rise to frustration will tend to provide a low dimensional driving with much wider access to the plethora of configurational degrees freedom. The configuration space may be huge, but the interactions determine how much of it can be accessed by the driving. 

\begin{textbox}[h]\section{Implications of NP on the $t$-graph topology}
The presence of NP implies: (i) under oscillatory shear, a limit cycle is reached after at most one driving cycle \cite{sethna1993hysteresis, munganterzi2018}, (ii) SCCs are $\ell RPM$ loop-graphs $(X,Y)$ and can therefore be exited only via $X$ or $Y$ \cite{munganterzi2018,terzi2020state}, and (iii) these exits happen when $\varepsilon > \varepsilon^+[Y]$ or $\varepsilon < \varepsilon^-[X]$. From (iii) it follows that if $A$ belongs to the SCC $(X,Y)$, any driving protocol $\varepsilon(t)$ such that $\varepsilon^-[X] < \varepsilon(t) < \varepsilon^+[Y]$  will confine the system to $(X,Y)$. 

Moreover, (i) via (ii) implies that an SCC $(X,Y)$ that is left via its upper exit $Y$ must lead to an SCC $(X',Y')$ that is more trapping in the forward and reverse direction, i.e. $\varepsilon^+[Y'] > \varepsilon^+[Y]$ and $\varepsilon^-[X'] \le  \varepsilon^-[X]$. Likewise, an exit from $(X,Y)$ via $X$ leads also to more trapping SCCs. This is a necessary condition for (i) to hold. Thus NP imposes strong constraints on the transitions between SCCs by canalizing the driven systems into increasingly more trapping SCCs \cite{munganterzi2018}. 
\end{textbox}

\subsection{Organization of soft-spots in determining flow properties}

\begin{figure}[t!]
\begin{center}
\includegraphics[width=\columnwidth]{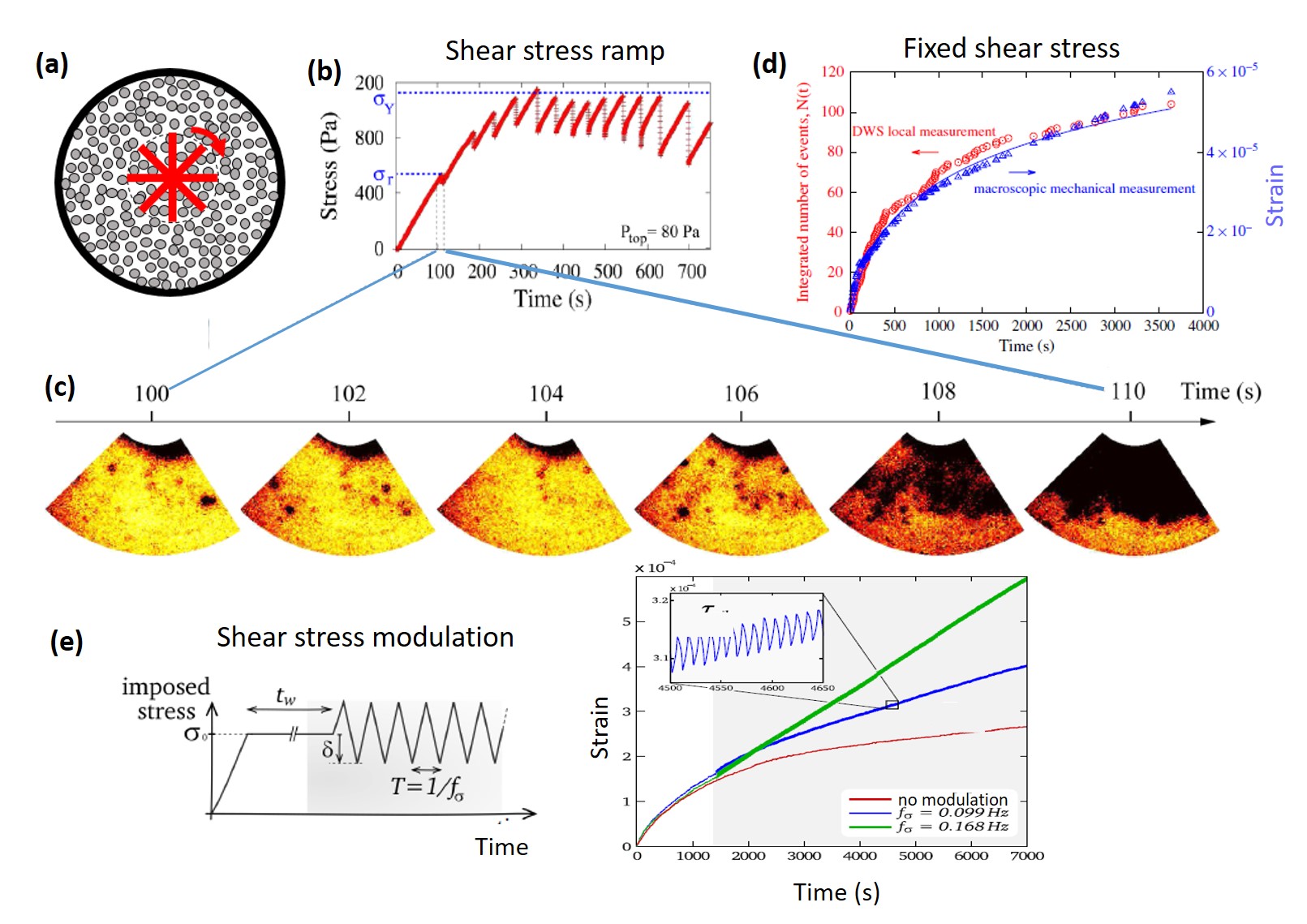}
\caption{{\bf Mesoscale emergence of ``soft spots" in a sheared granular packing. } {\bf (a)} Sketch (top view) of a shear cell containing glass beads vertically confined by a transparent load. Shear can be applied using a central vane (in red) to which a controlled torque is applied and the angular rotation measured. {\bf (b)} Shear stress evolution at constant external stress rate (stress ramp) showing a sequence of sudden stress drops. {\bf (c)} Direct visualization, during the charging process of ``hot spots", i.e. loci of large deformation spanning about 5-10 grain sizes,  using an optical Diffusing Wave Spectroscopy (DWS) technique. The darker the color, the larger the deformation. The stress drop at $\sigma_r$ corresponds to the spatial accumulation of ``soft spots" around a large shear band zone. {\bf (d)} Under constant shear stress, the logarithmic creep (in blue) quantitatively corresponds to the cumulated value $N(t)$ (in red) of all visualized ``soft spot" events, thus revealing the material support of an internal slowly varying variable interpreted as the ``fluidity" parameter in soft glassy rheology models.{\bf (e)} Under external stress modulation around a constant value, the logarithmic creep becomes linear due to the long-term accumulation of irreversible events materialized by a constant production rate of ``soft spots", corresponding equivalently to a dynamical irreversibility of the fluidity variable. Figure adapted from refs~\cite{Amon2012} and \cite{Pons2015}.}       

\label{fig:Granular_Eric}
\end{center}
\end{figure}

Although our discussion so far has been in the context of plastic rearrangements in glasses, drawing largely on computer simulation results, the idea of soft spots and their restructuring as {\it internal} variables that determine flow behavior encompasses a variety of soft solids, such as confined emulsions~\cite{Goyon2008}, colloidal glasses~\cite{Schall2007}, bubble rafts~\cite{Katgert2010}, foams~\cite{Kabla2003} or granular materials~\cite{Amon2012}. Within the energy landscape based framework of 
``soft glassy rheology''~\cite{Sollich1997,Derec2001}, the rate of transitions between energy states (equivalent to nodes in the t-graphs in the discussion above), often termed ``fluidity", plays a central role. A relationship between fluidity and emergent localized structures was proposed by Goyon et al.~\cite{Goyon2008} to provide a physical interpretation of confined emulsion experiments. It was directly visualized in granular matter~\cite{Amon2012}, where the spatiotemporal dynamics of localized deformations, equivalent to the ``soft spots" described above, (see Fig.~\ref{fig:Granular_Eric}) were linked to a fluidity variable that enters a simple ``soft glassy rheology'' model~\cite{Derec2001}. Creep dynamics and the suppression of the fluidization threshold by mechanical agitation in granular packings have also been interpreted in terms of the soft spot dynamics and its relation to fluidity~\cite{Pons2015,Pons2016}.

\subsection{Excursion: The Preisach model, return-point memory and mapping histories}
\label{subsec:preisach}

\begin{figure}[t!]
\begin{center}
\includegraphics[width=\textwidth]{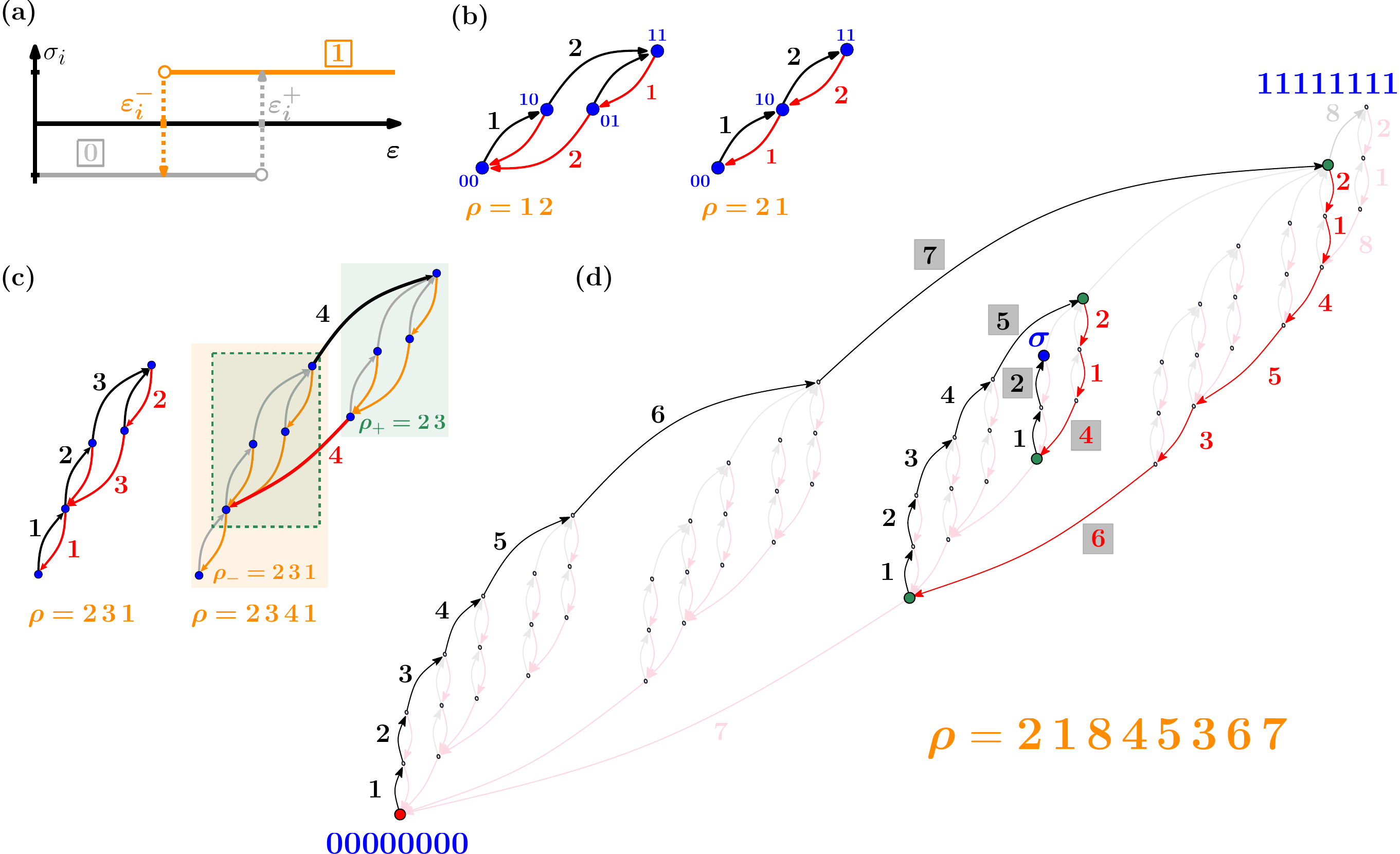}
\caption{{\bf The Preisach model.} {\bf (a)} A Preisach hysteron $i$ as a two-state system with hysteretic transitions between states {\tt 0} and {\tt 1} at switching fields $\varepsilon^\pm_i$. {\bf (b)} The two possible main hysteresis loop-graphs of a system of $N = 2$ independent hysterons. Such graphs are called Preisach graphs. Vertices are labeled by hysteron configurations. Labels next to  transitions indicate the  hysteron that changes state. Hysterons switch states independently and  we  label them according to the order in which they switch from state {\tt 0} to {\tt 1}, see black labels next to transitions from {\tt 00} to {\tt 11}. This leaves the order in which  hysterons switch back to {\tt 0} undetermined, which we prescribe by $\rho$. For $N = 2$ hysterons, the possible switch-back sequences are  $\rho = 1\,2$ and $\rho = 2\,1$, as indicated by the red labels next to the transitions from {\tt 11} to {\tt 00}.    {\bf (c)} The Preisach graphs of $\rho = 2\,3\,1$ and $\rho = 2\,3\,4\,1$. Due to loop return-point memory, the Preisach graph of $\rho = 2341$ can be described as a merger of the $\rho = 2\,3\,1$ graph (inside orange rectangle) and a certain subgraph (green rectangle inside) that is copied and connected to it by the two transitions involving the state changes of the fourth hysteron. {\bf (d)} The number of vertices of the Preisach graph of $\rho$ is equal to number of increasing subsequences contained in $\rho$. Each increasing subsequence can be seen as a deformation history that is applied to ${\tt 00\ldots 0}$, mapping it thereby to a distinct vertex of the Preisach graph. This is illustrated for the case $\rho = 2\,1\,8\,4\,5\,3\,6\,7$ and the increasing subsequence $2, 4, 5, 6, 7$, whose deformation path leading to $\bm{\sigma}$ is shown by the $\Up$- and $\Dn$-transitions higlighted in black and red, respectively.  
}
\label{fig:preisach}
\end{center}
\end{figure}

The system of persistent soft-spots that emerges from mechanical annealing emulates a collection of {\em hysterons}, i.e. two-state systems with hysteretic state-changes. In this excursion we assume that such a soft-spot system has been established by annealing,  and we will study how these encode memories of the driving histories. This leads us to the Preisach model, which was introduced  by F. Preisach in 1935 to understand hysteresis in magnets \cite{Preisach1935}. 
The model describes the hysteretic switching of a collection of hysterons but ignores interactions between these\footnote{Systems with interacting hysterons were initially introduced in \cite{hovorka2005onset} and are receiving renewed interest \cite{bense2021complex, keim2021multiperiodic, lindeman2021multiple, van2021profusion, szulc2022cooperative,lindeman2023isolating, muhaxheri2024bifurcations, liu2024controlled, sirote2024emergent}. They were reviewed recently in  \cite{paulsen2024mechanical}.}. Still,  it is a useful model illustrating clearly how memory of deformation history is encoded into internal states  \cite{jules2022}. Our excursion is based on \cite{terzi2020state,ferrari2022liss}.

The Preisach hysteron is defined by two switching thresholds $\varepsilon^\pm$, such that when $\epsilon < \varepsilon^+$, the hysteron can be in state ${\tt 1}$, whereas it can be in state ${\tt 0}$ when $\epsilon > \varepsilon^-$, as illustrated in Fig.~\ref{fig:preisach}(a). Hysteresis occurs when $\varepsilon^- < \varepsilon^+$. Between $\varepsilon^-$ and $\varepsilon^+$, the hysteron can be in either of its states, and the actual state chosen will depend on the driving history.  The Preisach model exhibits return-point memory \cite{Barker1983}, thus its $t$-graph obeys $\ell$RPM.   

In the discrete Preisach model, we consider a set of $N$ hysterons,  each with a pair of distinct switching thresholds and such that $\varepsilon_i^- < \varepsilon_i^+$. It is assumed that each hysteron couples independently to the external drive $\varepsilon$. 
Thus there are $2^N$, possible hysteron configurations. We call the two configurations in which all hysterons are either in state ${\tt 0}$ or ${\tt 1}$, the saturated configurations and denote them by ${\bf  0}$ and ${\bf  1}$, respectively. In the following, it will be sometimes more convenient to identify hysteron configurations by specifying the set $I$ of hysterons that are in state ${\tt 1}$, so that 
${\bf  0} \leftrightarrow \emptyset$, while ${\bf  1}  \leftrightarrow \{1, 2, \ldots, N\}$.

A hysteron configuration $I$ is said to be stable at a drive $\varepsilon$, if the state assignment of each hysteron is   compatible with its switching fields: for each $j \in I$ we require $\varepsilon^-_j < \varepsilon$, while for each $i \in I^c$ the condition $ \varepsilon < \varepsilon^+_i$ must hold, with $I^c$ being the set-complement of $I$. A necessary and sufficient condition for all of these inequalities to be simultaneously satisfied, and thus for  $I$ to be stable, is given by  
\begin{equation}
\varepsilon^-[I] \equiv \max_{j \in I} \, \varepsilon^-_j 
< \min_{i \in I^c} \, \varepsilon^+_i \equiv \varepsilon^+[I].
\label{eqn:preisach-stable}
\end{equation}
For the two saturated states, we shall assume that $\varepsilon^-[\emptyset] = -\infty$, while, $\varepsilon^+[\{1, 2, \ldots, N\}] = \infty$. 
The mesostates  of the Preisach model are its stable states, i.e. those sets $I$ that satisfy the inequalities of \eqref{eqn:preisach-stable}. The stability range of a mesostate $I$ is then given by $\varepsilon^-[I] < \varepsilon < \varepsilon^+[I]$.  

As is apparent from \eqref{eqn:preisach-stable}, to determine whether a configuration of hysterons $I$ is a mesostate or not, it suffices to know the rank ordering of the set of switching fields $\varepsilon^\pm_i$. A knowledge of their actual values is not necessary.  Since hysterons couple independently, without loss of generality we assume that 
\begin{equation}
\varepsilon^+_1 < \varepsilon^+_2 < \ldots < \varepsilon^+_N.
\end{equation}
and let $\rho = (\rho_1, \rho_2, \ldots, \rho_N)$ be the permutation of $\{1, 2, \ldots, N\}$, that orders the lower switching fields from largest to smallest as 
\begin{equation}
\varepsilon^-_{\rho_1} > \varepsilon^-_{\rho_2} < \ldots < \varepsilon^-_{\rho_N}.
\end{equation}
The permutation $\rho$ prescribes the order in which the $N$ hysterons switch their states back to {\tt 0}. For example, starting from the configuration {\tt 11\ldots 1}, first hysteron $\rho_1$, then $\rho_2$ etc. switches. 
Details on the role of the ordering of switching fields can be found in \cite{terzi2020state, daskrugmungan2022}. 
Here we will be interested in the  main hysteresis loop and its sub-cycles, i.e. the loop graph bounded by the two saturated states ${\bf 0}$ and ${\bf 1}$. We call this loop graph the Preisach graph. We show next that the topology of the Preisach graph is uniquely determined by $\rho$.  

To do so, we consider first the transitions out of a mesostate $I$.  Denote by $i^\pm[I]$ the sites at which minimum and maximum of the inequalities in \eqref{eqn:preisach-stable} are attained, so that $ \varepsilon^-_{i^-} = \varepsilon^-[I]$ and $ \varepsilon^+_{i^+} = \varepsilon^+[I]$. These are the least stable sites, i.e. the sites that will lose stability first when $\varepsilon$ is reduced to $\varepsilon^-[I]$ or increased to $\varepsilon^+[I]$. Consider the case when $\varepsilon = \varepsilon^+[I]$, so that $i^+[I]$ becomes unstable. Denote by $J = I \cup \{ i^+[I]\}$, the hysteron configuration obtained by changing the hysteron state of $i^+$ from ${\tt 0}$ to ${\tt 1}$. It is readily shown \cite{terzi2020state} that (i) $J$ is a stable configuration and (ii) $J$ is stable at $\varepsilon^+[I]$, i.e. the inequality $ \varepsilon^-[J] < \varepsilon^+[I] < \varepsilon^+[J]$ holds. An analogous result holds for the transition at $\varepsilon = \varepsilon^-[I]$ and these together define the $\Up$- and $\Dn$-transitions out of  mesostate $I$. 
An immediate consequence of the above observations is that the Preisach model does not exhibit avalanches, since state changes under $\Dn$ or $\Up$ involve a single hysteron flip only. 

The Preisach graphs for $N = 2$ can be worked out readily by 
starting in the lower saturated state ${\tt 00}$ and following the $\Dn$- and $\Up$-transitions, as defined above. The two $N=2$ Preisach graphs, corresponding to $\rho = 1\,2$ and $\rho = 2\,1$, are shown in Fig.~\ref{fig:preisach}(b). The hysteron configurations are indicated next to each mesostate. The label besides a transition indicates the hysteron that changes its state. 

The left part of Fig.~\ref{fig:preisach}(c) shows the Preisach graph for $N = 3$ and $\rho = 2\,3\,1$. Consider next the $N = 4$ Preisach graph with $\rho = 2\,3\,4\,1$. We may think of this as augmenting the  $\rho = 2\,3\,1$ system by a fourth hysteron as follows.  Starting again from the lower saturated state $\emptyset$, as long as we do not cause a state change of the $4$th hysteron to ${\tt 1}$, the possible transitions are  those given by the Preisach graph of $\rho = 2\,3\,1 \equiv \rho_-$. This is the sub-graph contained inside the orange rectangle. Clearly, by our ordering of the upper switching fields, the $4$th hysteron can only switch to ${\tt 1}$ after all other hysterons are in state ${\tt 1}$, i.e. when we are in the state ${\tt 1110}$, leading then to the upper saturated state ${\tt 1111}$. Once there, we can now reduce and increase the driving as long as hysteron $4$ does not revert its state to ${\tt 0}$. Since $\rho = 2\,3\,4\,1$, this means that (i) we can only change the states of hysterons $2$ and $3$, and (ii) $4$ will return to state ${\tt 0}$ only after hysterons $2$ and $3$ are in state {\tt 0}, i.e. when reaching {\tt 1001}, so that after switching back $4$, the hysteron configuration will be ${\tt 1000}$. In fact, the loop graph bounded by ${\tt 1001}$ and ${\tt 1111}$ is generated $\rho_+ = 2\,3$, which is  isomorphic to the Preisach graph with $\rho = 1\,2$ (rectangles shaded in green). Note that this loop graph is ``copied" from the corresponding sub-graph generated by $\rho_- = 2\,3\,1$ (green box inside the orange rectangle). From our discussion it is clear that (i) the two permutations $\rho_-$ and $\rho_+$ describe the two sub-graphs and (ii) that these two Preisach graphs are joined into the Preisach graph of $\rho$. The two permuations $\rho_-$ and $\rho_+$ can be generated from $\rho$ as follows. The permutation $\rho_-$ is obtained from $\rho$ upon removal of the element $N$, while $\rho_+$ is the permutation obtained by removing from $\rho$ the element $N$ as well as all other elements that come after it. 

Denote by $P(\rho)$ the number of vertices of the Preisach graph that is generated by $\rho$. The construction of the Preisach graph of $\rho$ from the Preisach graphs of the permutations $\rho_\pm$ described above, implies that 
\begin{equation}
P(\rho) = P(\rho_-) + P(\rho_+).
\end{equation}
Treating $\rho$ as a sequence, define an {\em increasing subsequence} as a subsequence of elements of $\rho$ that is strictly increasing.
\begin{marginnote}[]
\entry{Increasing Subsequence}{In $\rho = 1\,2$, the increasing subsequences are, the empty sequence $()$,  $(1)$, $(2)$ and $(1,2)$. For $\rho = 21$, there are only three.  
}
\end{marginnote}
Denote by $\mathcal{N}(\rho)$ the number of increasing subsequences contained in $\rho$. It turns out that $P(\rho) = \mathcal{N}(\rho)$, i.e. the Preisach graph generated by $\rho$ has as many vertices as their are increasing subsequences of $\rho$. 
To see this,  let $\rho$ be a permutation of $N > 1$ elements. The increasing subsequences of $\rho$ either contain $N$ and hence end in $N$, or they do not. The number of the former kind is $\mathcal{N}(\rho_+)$, while that of the latter is given by $\mathcal{N}(\rho_-)$, so that $\mathcal{N}(\rho) = \mathcal{N}(\rho_-) + \mathcal{N}(\rho_+)$. Given that $\mathcal{N}(1) = P(1) = 2$, and proceeding inductively, it follows that $\mathcal{N}(\rho)= P(\rho)$.

The equality of the number of increasing subsequences and vertices of the corresponding Preisach graph is not a coincidence. As was shown in \cite{ferrari2022liss}, there is a one to one mapping between these two sets that moreover highlights the role of return-point memory underlying the topology of the Preisach graph. Taking again the lower saturated state as a reference state and assigning it the empty subsequence, we assign to each increasing subsequence of $\rho$ a mesostate as follows. We regard the corresponding increasing subsequence as a driving history that, when applied to ${\tt 00\ldots 0}$, leads to the corresponding state it is mapped to. This is illustrated in Fig.~\ref{fig:preisach}(d) for the Preisach graph generated by $\rho = 2\,1\,8\,4\,5\,3\,6\,7$ and the increasing subsequence $(2,4, 5, 6, 7)$. Reading this subsequence from right to left, and starting from ${\tt 00000000}$, the sequence of applied drives is as follows: we first increase the driving until the $7$th hysteron changes its state from ${\tt 0}$ to ${\tt 1}$. Next, we decrease the driving until the $6$th hysteron reverts its state to ${\tt 0}$. This is then followed by an increase of the driving until hysteron $5$ is in state ${\tt 1}$, a subsequent decrease until hysteron $4$ is at ${\tt 0}$ and a final increase until hysteron $2$ is in state ${\tt 1}$. We assign the mesostate $\bm{\sigma} = {\tt 11101010}$ reached in this manner to the driving history $(2,4, 5, 6, 7)$. The proof of this one-to-one correspondence 
and its connection to return-point memory is given in \cite{ferrari2022liss}. 

We conclude with a few observations. The Preisach graph is an SCC and the mapping establishes a correspondence between driving histories and its mesostates. The possibility of such a construction is due to the nested hierarchy of hysteresis cycles which is a consequence of $\ell$RPM. It furnishes a filtering of the deformation histories, playing the role of a decision tree. In fact, the hierarchical nesting of the graph-loops can be represented in terms of a tree \cite{munganterzi2018, terzi2020state}. To see this, refer again to the right panel of Fig.~\ref{fig:preisach}(c). We can view this as a decomposition of the Preisach graph of $\rho = 2\,3\,4\,1$ into the two Preisach graphs generated by $\rho_- = 2\,3\,1$ and $\rho_+ = 2\,3$ and connected to each other by the transitions of the $4$th hysteron. The $\rho_- = 2\,3\,1$ graph consists out of all mesostates in which hysteron $4$ is in state ${\tt 0}$, while its state in the $\rho_+ = 2\,3$ graph is ${\tt 1}$. Since, each of these sub-graphs is a Preisach graph itself, the same decomposition into two sub-graphs can be performed on these as well. This decomposition is continued until all Preisach graphs contain single vertices, they are the Preisach graph generated by the empty permutation, i.e. $N = 0$.       

The Preisach model, despite its limitations, demonstrates how a system of hysteretic bi-stable switches is capable of recognizing patterns of environmental histories, mapping such information into states, by organizing the states and the transitions between theses into a $t$-graph with nested cycles. Thus the hysteron configurations and the transitions between them establish a representation of the history of the  environment that drives it. We have seen that such persistent hysteron systems emerge as a result of driving by oscillatory shear. 

\section{Self-organization in a fluctuating environment}
\label{sec:self-organization}

In the previous section we have described the self-organization of an amorphous solid when subject to oscillatory shear. The driving led to the emergence of a persistent soft-spot system that established the mechanically reversible behavior. These observations lead to the question, whether driving by cyclic shear is necessary for establishing a persistent system of mechanical instabilities, and more specifically, whether such self-organization can be achieved also by random driving protocols. In this section we consider the response of an amorphous solid when subjected to driving by either (i) a randomly fluctuating external shear or (ii) active particles, where each particle experiences also  Langevin forces. In both cases it is found that the response self-organizes into a state of high mechanical reversibility.    

\subsection{Memory formation under random driving} 
\label{subsec:randomdr}

We first ask whether a random deformation protocol confined to a strain range $\pm \varepsilon_T$ suffices to establish a persistent interacting soft-spot system \cite{mungan2022comm}. 
That this is indeed the case was shown recently in simulations of a mesoscale elastoplastic model of an amorphous solid with a {\em quenched} disorder landscape, called the QMEP model \cite{mungan2024self}.  The QMEP model reproduces qualitatively and quantitatively many features of the AQS phenomenology, including yielding, the irreversibility transition, topological properties of $t$-graphs extracted from it \cite{kumar2022}, as well as memory formation under cyclic shearing \cite{kumar2024self}, cf. Fig.~\ref{fig:amorphous-intro}h.  The sidebar summarizes key features of the QMEP model. 

\begin{figure}[t!]
\begin{center}
\includegraphics[width=\textwidth]{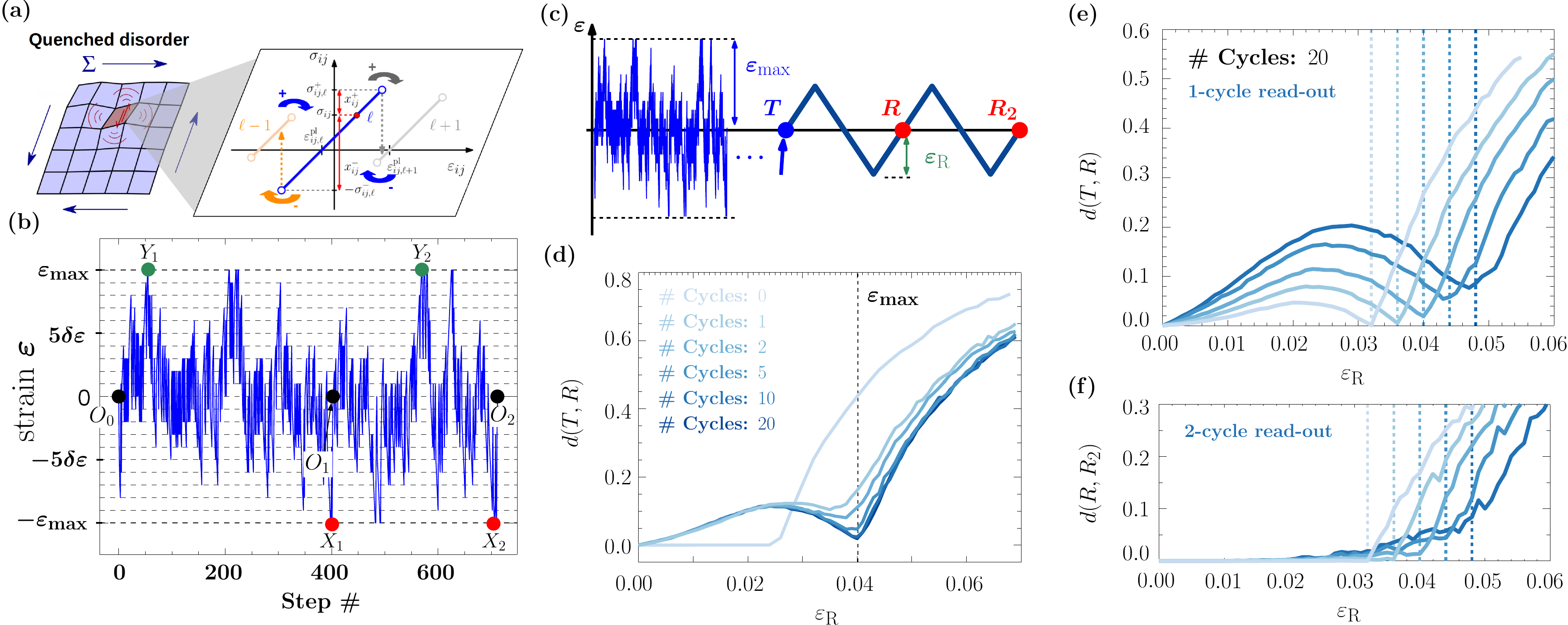}
\caption{{\bf Self-organization and memory in a mesoscale model of an amorphous solid subject to random shear loading (panels ordered column-wise).} {\bf (a)} (Left) 2d lattice of mesoscale cells. (Right) The response of each cell $(i,j)$ to an external stress $\Sigma$ follows a sequence of local elastic branches, labeled $\ldots, \ell-1, \ell,\ell + 1, \ldots$, which are each limited by stress thresholds in the forward $\sigma_{ij,\ell}^+$ and reverse $\sigma_{ij,\ell}^-$ directions of shear. 
A mesostate $A$ corresponds to the matrix $\ell_{ij}$, indicating the local branch index of each cell. {\bf (b)} Sample history by application of random strains $\varepsilon$ to an initial glass $O_0$. The strain protocol is a random walk along the strain axis with fixed steps $\delta \varepsilon$ and reflecting boundaries at $\pm \varepsilon_{\rm max}$. 
The boundary reached first defines the sense of driving and is called the first boundary. A cycle is defined in terms of the first passage times to go from zero strain to the first boundary, then to the opposite boundary and to subsequently return to zero strain, like the cycle $O_0 \to Y_1 \to X_1 \to O_1$ shown. 
{\bf (c)} Definition of the  (deterministic) 1- and 2-cycle read-out protocols. In the latter, a second read-out cycle of the same amplitude is applied to $T$. The configurations reached at the end of the first and second read-out cycle are $R$ and $R_2$. 
{\bf (d)} Results of 1-cycle read-out applied to the trained state $T$ reached after various numbers of random training cycles, cf. Fig.~\ref{fig:amorphous-intro}(f).  {\bf (e)} 1-cycle read-outs from glasses subjected to 20 random cycles at different  amplitudes $\varepsilon_{\rm max}$, as indicated by dashed  verticals.
{\bf (f)} 2-cycle read-out of $d(R,R_2)$, cf. {\bf (c)}. 
Panels (a) -- (f) adapted from \cite{kumar2024self, mungan2024self}.
} 
\label{fig:random-driving}
\end{center}
\end{figure}

\begin{textbox}[h]\section{The QMEP mesoscale model of an amorphous solid}
A $2d$ amorphous solid is coarse-grained into an $N\times N$ lattice of mesoscale cells $(i,j)$, see  Fig.~\ref{fig:random-driving}(a). With each cell we associate a stack of local elastic branches, which are indexed by integers $\ell_{ij}$ and capture the local stress strain response under elastic deformations. Each local elastic branch $\ell$ is characterized by quenched random variables: the local plastic strain $\varepsilon^{\rm pl}_{ij,\ell}$ and stress thresholds $\sigma^{\pm}_{ij,\ell}$, cf. Fig.~\ref{fig:random-driving}(a). Once the total stress on a local elastic branch exceeds its stress threshold, a transition to a neighboring branch occurs, accompanied by a redistribution of internal stresses which may cause avalanches. 

The mechanical properties of a global elastic branch $A$ are determined via the local branch properites by specifying for each cell $(i,j)$ its branch index $\ell_{ij}[A]$, and thereby define the mesostate $A$. The limits $\varepsilon^\pm[A]$ of global elastic branches are determined by the strain at which the first cell reaches it local threshold. This is very similar to how instabilities are triggered in the Preisach model of Section \ref{subsec:preisach}. Note that while in the QMEP model transitions between local branches are reversible, those between global branches are generally not, since cells triggering the $\Up$- and $\Dn$-transitions out of it  will be different.

\end{textbox}

We take our random deformation protocol to be an unbiased random walk along the external strain axis with reflecting boundaries at shear strains $\pm \varepsilon_{\rm max}$. Starting at the origin $\varepsilon = 0$, our random driving consists of making strain steps of fixed size  $\delta \varepsilon = \varepsilon_{\rm max}/ M $, so that it takes at least $M$ steps for the walker to reach one of the reflecting boundaries from the origin. 
In analogy to driving by {\em deterministic} cyclic shear, we define a {\em random} shear cycle by first passages from one boundary to the other and back \cite{mungan2024self}. 
Fig.~\ref{fig:random-driving}(b) shows a typical deformation path with the freshly 
prepared glass $O_0$ at zero applied strain, and the first cycle leading first to $+\varepsilon_{\rm max}$ at $Y_1$ then $-\varepsilon_{\rm max}$ at $X_1$, and a  return to zero strain at  $O_1$. Further details of the random walk implementation and cycle definition are given in \cite{mungan2024self}. 

We apply $\mathcal{N}$ random cycles to a fresh glass $O_0$ at the end of which we reach the trained mesostate $T$. Read-outs are performed  by subjecting copies of $T$ to single cycles of {\em deterministic} oscillatory cyclic shear at amplitudes $\varepsilon_R$, cf. Fig.~\ref{fig:random-driving}(c). Fig.~\ref{fig:random-driving}(d) shows the result for read-outs after various numbers of random training cycles.
The distance $d(T,R)$ between trained and read-out mesostates is determined from their branch configurations $\ell_{ij}[T]$ and $\ell_{ij}[R]$ as the 
 fraction of cells $(i,j)$ for which $\ell_{ij}[T] \ne \ell_{ij}[R]$, i.e. the fraction of cells that  at the end of the read-out cycle have not returned to their local branches at $T$.
Already after a few random training cycles, a cusp of $d(T,R)$ develops at $\varepsilon_R \approx \varepsilon_{\rm max}$, getting more pronounced with increasing cycles. Fig.~\ref{fig:random-driving}(e) shows read-outs from glasses trained for 20 random cycles with different strain ranges (vertical dashed lines), demonstrating 
that a memory of $\varepsilon_{\rm max}$ is encoded. 

Fig.~\ref{fig:random-driving}(f) shows the result of applying two read-out cycles to $T$ and comparing the states $R$ and $R_2$ reached at the end of the first and second cycle, as illustrated in Fig.~\ref{fig:random-driving}(c). Referring to point (i) of the sidebar describing the consequences of NP, if NP were to hold, then $d(R,R_2) = 0$, {\em for any} amplitude $\varepsilon_R$ and state $T$ to which the two-cycle read-out is applied, i.e. regardless of whether $T$ is a result of some prior training or not~\cite{middleton1992asymptotic, sethna1993hysteresis, munganterzi2018}. What we see instead is that $d(R,R_2)$ remains close to zero only as long as $\varepsilon_R \lesssim \varepsilon_{\rm max}$, beyond which it starts to rise quickly. Thus while the first driving cycles moves the configuration away from $T$ and towards $R$, the second cycle returns it close to $R$. When restricted to $\varepsilon_{\rm R} \lesssim \varepsilon_{\rm max}$, the trained glasses exhibit a highly mechanically reversible response, while for read-out strains larger than $\varepsilon_{\rm max}$, reversibility is rapidly lost. The application of an external random driving  thus suffices to imprint a mechanical memory on the disordered solid. Qualitatively similar results were recently obtained from molecular dynamics simulations of an amorphous solid \cite{chatterjee2025randomlydriven}. 

\subsection{Activity-induced annealing of an amorphous solid}

Another form of driving that has recently been actively investigated, bearing strong analogies to driving by applied shear, as well as incorporating elements of spatial and temporal fluctuations, is driving by active forces. There has been intense interest in investigating  the dynamics and steady states of self-propelled particles \cite{SRARCMP2010,ActiveMatterRMP2013}, with emphasis on modeling biological systems across length scales, from bird flocks to the cytoskeleton and other subcellular structures. Some of these contexts, such as modeling confluent tissue \cite{BiPRX2016} have motivated investigation of active glasses (dense assemblies of particles propelled by active forces), with the transition from jammed to fluidized states being of particular interest \cite{Xu2018,mandal2020extreme,MorsePNAS21,Agoritsas_2021,During2021,sharma2025activity,goswamy2025}. The role of active forces in the context of memory has also received attention in recent times \cite{behera2023enhanced-memory,du2025memory-information,agoritsas2024memory,sharma2025activity,agnish2024}. Memory retrieval in the Hopfield model of associative memory, and related systems, in the presence of active noise has been analyzed recently, to show that an active component in the noise leads to an enhancement of memory capacity \cite{behera2023enhanced-memory,du2025memory-information,agnish2024}. An appealing connection between the presence of active noise, and memory enhancement through the process of Hebbian ``unlearning'' or ``dreaming'' has been discussed in \cite{agnish2024}.  These works address the role of active driving in memory retrieval. In addition, memory formation through the application of active driving, in direct analogy with memory formation under cyclic or random shear discussed above, have also been addressed recently \cite{sharma2025activity,agoritsas2024memory}. In \cite{sharma2025activity}, memory formation under driving by active forces, with a finite persistence time for the active propulsion direction was studied. The ``read'' protocol involves the application of active driving of different magnitudes for a short period of time. Rather than the displacements of particles, the additional annealing, in the form of the magnitude of the change in energy, is monitored. The magnitude of the energy change exhibits a minimum at the training magnitude of the active force, and thus a memory. This memory effect bears resemblance both to the memory formation under external shear discussed here, and memory in thermal glassy systems (including spin glasses and molecular glasses) of the temperature at which the glass is aged \cite{jonason1998,Yardimci2003}. Significantly for the present discussion, solids driven by active forces presents another example of memory formation under fluctuating driving conditions, which also has significant biological relevance. 

\section{Self-organization and learning: From non-living to living systems} 
\label{sec:to_live_or_not}
 
In the preceding sections, we have described the mechanical annealing of a sheared amorphous solid when subjected to oscillatory and random protocols of driving. Under these conditions, the system self-organizes its response by the triggering of spatially localized instabilities, the soft-spots, which eventually leads to a steady-state, whose mechanical response is governed by a collection of persistent soft-spots. The soft-spots are multi-stable and transitions between their local states are hysteretic. The persistence of the soft-spots means that their instabilities can be repeatedly triggered and in turn gives rise to the high-degree of mechanical reversibility. The imposed driving shapes the steady-state mechanical response of the amorphous solid and in this way imprints a memory. Such memories can encode patterns of environmental histories. Along the way, we have seen that the $t$-graph description provides a natural way to track the evolution of mechanical instabilities and to probe the emergence of mechanical reversibility.

\subsection{Soft-spot transitions, slow-manifolds and soft-modes}

The $t$-graph description, by focusing on the transitions between mesostates, has coarse-grained the AQS dynamics into a pair of maps $\Up$ and $\Dn$ describing the transitions between mesostates under increase and decrease of the deformation parameter, cf. Fig.~\ref{fig:tgraph-intro}(a) and (b). As we have seen, dynamical features such as memory formation and reversibility manifest themselves at the level of the $t$-graph. However, the graph-topology depends on the underlying microscopic properties, namely the relaxation pathways that are followed when mesostates become unstable, giving rise to plastic events. These relaxations are governed by the collection of multistable and spatially localized soft-spots. Their state change induces a long-range redistribution of local elastic stresses that can create, disable or alter the switching behavior of other soft-spots. These interactions govern the transition pathways of plastic events. The dynamics of soft-spots thus forms a description of the amorphous solid at a mesoscopic scale, an intermediate scale that is coarser than the particle configurations, yet finer than the coarse-grained $t$-graph description. This is the idea behind introducing mesoscale models like the QMEP model, described in section \ref{subsec:randomdr}.

The dynamics of soft-spots is closely related to the quadrupolar nature of the stress-redistributions, which gives rise to frustration. There is an intuitive way to see why such a form of stress-distribution can lead to spatially localized instabilities. Consider first a stress redistribution of the kind where  the local stress-drop of a soft-spot that yielded would be redistributed such that only additional loads are placed on the rest of the system. Ferromagnetic interactions in magnets are of this type. It is also what happens when parts of an elastic interface depin and the load that is relieved is passed on to the other parts of the interface. At a macroscopic length scale, this corresponds to a diffusion of local stresses. In the long run, such diffusive processes tend towards an equalization of stresses across the system. Owing to its quadrupolar nature, stress redistributions under shear have also a component that resembles negative diffusion, i.e. diffusion with a negative diffusion coefficients. This implies that at some parts of the sample local stresses, rather than diffusing away, may become more concentrated. Evidence for such behaviour and related phenomenology has been observed in a wide range of amorphous solids \cite{nicolas2018deformation}, including granular materials, as those described in 
Fig.~\ref{fig:Granular_Eric}, involving frictional interactions \cite{Amon2012,Pons2015,LeBouil2014}.

Soft-spots are multistable, transiting from one state to another involving localized rearrangements. In the case of a persistent soft-spot, such transitions are also reversible. The rearrangement of local particle configurations follows a slow manifold which is established in terms of soft-modes. A very accessible exposition of these concepts has been given in  \cite{russo2025softmodes}. 
In the context of amorphous solids and interparticle interactions,  we can think of the configuration space characterised by  soft mode transitions as low-dimensional valleys in a high-dimensional potential energy landscape, towards which the system evolves under training. As a result, we may expect the system configuration to move along such valleys in a similar fashion regardless of the type of external perturbation. Thus local particle rearrangements under different types of loading will be similar.  This is why and how a collection of soft-spots prepared by some annealing protocol can respond rather similarly when driven by AQS shear, low temperature fluctuations \cite{lerbinger2022supercooled} or active driving. If in each case the same soft-spots are activated, the final mechanical response will be the same.  

It is believed that soft-modes underlie diverse relaxation mechanisms when a biological equilibrium state is perturbed by environmental or internal changes. In each of these cases the perturbations may project down on the same soft modes and hence follow a common response pathway \cite{russo2025softmodes}. We will discuss this point in Section \ref{sec:bio}. 

\subsection{Mirroring of the dynamics of the system and its environment establishes a representation of the latter by the former}

The amorphous solid subject to random driving is a prototype for a system interacting with a fluctuating environment. 
The self-organization leads to a mechanically reversible steady-state in which the subsequent response of the system is correlated with that of the driving. As we have seen in the case of the Preisach model, the triggering sequence of instabilities permits one to infer features of the dynamics of the environment, e.g. Fig.~\ref{fig:preisach}(d). In this way, the soft-spot states furnish a digital encoding of their environmental history. These correlation persists as long as the range of the fluctuations of the driving does not change appreciably. 

\begin{textbox}[h]\section{Pattern recognition and machine learning}
A classical pattern recognition (PR) task is the recognition of hand-written digits representing $D = \{0, 1, . . . , 9 \}$. 
Samples provided as $N \times N$ pixel images, can be described by vectors $\bm{x}$ in an $N^2$ dimensional space $\mathcal{X}$ whose components represent the intensity of each pixel. 
In both {\em supervised} and {\em unsupervised learning}, the task is to determine a classifier, i.e. a map $y$ from $\mathcal{X}$ to $D$, assigning to a given hand-written digit image $\bm{x}$ the label $d \in D$ it represents. 
In supervised learning, $y$ is a parametrized function,  “learned” by finding the parameters minimizing the classification error in a {\em training set} of images  and their {\em ground truth} labels,  
$(\bm{x}_i,d_i)_{i = 1}^\mathcal{N}$. 
The estimation of $y$ furnishes the classification regions $R_d = \{ \bm{x} \in \mathcal{X} : y(\bm{x}) = d \}$. 
In unsupervised learning, there are only input samples $\bm{x}_i$ but no labels. In algorithms like {\em $k$-means clustering}, the learning goal is determining from the input data cluster centers in $\mathcal{X}$ for each  digit $d$. New inputs $\bm{x}$ are classified as the digit $d$ whose cluster center is nearest. The cluster centers provide a {\em quantization} of $\mathcal{X}$, furnishing a compressed representation of digits. A good introduction to PR and machine learning is \cite{bishop2006pattern}.
\end{textbox}

In fact, we can think of the system's interactions with its  fluctuating environment as having established an internal representation of its environment: the dynamics of the environment is mirrored by the dynamics of the system's instabilities that are triggered by it \cite{tagkopoulos2008predictive, libchaber2020walking}. 
Evidently, the process of self-organization has led to a capability which in the context of machine learning would be associated with pattern recognition and unsupervised learning of a representation, but without a pre-defined task and a generic optimization or training procedure to achieve it (see side bar). The correlation between the system's response at steady-state and the fluctuations of the environment that they mirror can be regarded as an elementary form of environmental sensing. 
Given that a model of a disordered system with minimal ingredients -- disorder, a built-in frustration in redistribution of external loads and negligible thermal effects -- is able to self-organize in a manner described above, one may wonder whether simple biological organisms lacking a nervous system, repurpose such  essentially ``for free" infrastructure when adapting to changing environments. 
The title of a paper that addresses such possibilities in a general setting summarizes this point rather well by asking whether {\em self-organization proposes what natural selection disposes}~\cite{batten2008visions}. In the following section we will address this question, by reviewing the biological evidence, as well as theoretical approaches that have been put forward. The sidebar on cell-motility and chemotaxis serves as a motivation. 

\begin{textbox}[h!]\section{Motivation: Cell-Motitility and Chemotaxis}
Bacteria can sense environmental gradients of nutrients, stresses or magnetic fields and control their motion accordingly. The resulting motion is called chemo-, duro and magnetotaxis, respectively. 
The cytoskeleton of eukaryotic cells can adjust its mechanical properties according to mechanical deformations exerted by its environment in an active manner  \cite{gralka2015biomechanics, adar2024ecmmemory}. In microbiology, the molecular and protein elements involved in the sensory circuitry of E. coli as well as their link to microorganism motility are well-studied \cite{phillips2020molecular}. From a functional point of view, bacterial chemotaxis can be viewed as a coupled network of bi-stable  elements \cite{Bray1998}. Each receptor is an input site for the ligand, which transmits information to the motor, the output being a possible change in swimming direction. The sensory elements carry essentially two activity states, whose switch point is controlled by a feedback loop stemming from a methylation process of the receptors that achieves memory, adaptation and coupling between the different receptors (see review by Lan et al.\cite{Lan2016}). This combination leads to the ability to dynamically compute chemical gradients with a high sensitivity that can only be understood as a cooperative switch effect involving tens of sites. 
\end{textbox}

\section{What can memory and self-organization in driven disordered systems teach us about how simple organisms adapt to changing environments? }
\label{sec:bio}

Sensing its environment by processing environmental cues and reacting to these, is essential for the survival of organisms in a changing environment and determines their evolutionary success.  The adaptation of even relatively simple organisms, such as bacteria, to changing environments must involve some form of information processing, memory and learning~\cite{casadesus2002memory, tagkopoulos2008predictive,hilker2016priming,marijuan2021molecularrecognition, gershman2021learning,Murugan_2021, timsit2021molecularbrain,vermeersch2022microbes,britto2025prbio,omer2025evolutionarylearning}.

\subsection{History-dependent behavior in simple organisms -- the biological evidence}

The issue of memory and learning in unicellular organisms has been revisited in two recent reviews  \cite{gershman2021learning,vermeersch2022microbes}.  As  summarized in Gershman et al. 
\cite{gershman2021learning}, the question whether unicellular organisms, which lack a nervous system, can learn, for example by exhibiting conditioned reflexes dates back to the beginnings of the last century, and is debated to this day \cite{sourjik2024consciousness}.

Part of the difficulty is that in vivo experiments which aim to demonstrate phenomena of memory formation and learning, require  controls for many confounding factors that could lead to false-positive results. This becomes more of an issue the more complex the organism is -- even when it is unicellular. 
There is increasing evidence that metabolic regulation of the cell exhibits history-dependent behavior (HDB), as discussed in 
the review by Vermersh et al. \cite{vermeersch2022microbes}, which is part of a special issue on non-genetic memory in microorganisms. Its focus is on HDB, which the authors take  as an operational definition of memory. While the review discusses mostly eukaryotic unicellular organisms and pathways of memory formation that are specific to these, it highlights the role of non-DNA based mechanisms of inheritance that are part and parcel of any unicellular organism that multiplies via cell division. 
\begin{marginnote}[]
\entry{Unicellular Eukaryotes and Prokaryotes}{Fungi and some algae are eukaryotes and are highly evolved. Bacteria are prokaryotes characterized by a simpler structure, e.g. lacking cellular organelles}. 
\end{marginnote}

Particularly relevant for bacteria is the perfect or imperfect cloning of its metabolic state as a result of division. This is how the protein contents of a cell, including those responsible for gene regulation, are passed on to subsequent generations. In this way, a stress-response pathway may remain active over many generations, even though the source of stress has subsided. Likewise, the concentration of any metabolite that is no longer actively synthesized will be successively diluted from generation to generation, thereby providing a mechanism of forgetting as the stimulus fades away over time \cite{lambert2014memory}.

This is one of the reasons why exploring phenomena of memory formation and learning in much simpler non-living settings,  such as that of driven disordered systems, is of interest. 
Such studies provide minimal model systems with a limited yet well-understood set of parameters that can be controlled. At the same time, these efforts enable us to uncover the necessary and sufficient conditions for such phenomena to emerge.  Gaining such insights and having simple paradigmatic models at one's disposal, should provide a sound perspective from which to re-assess the extent that simple organisms can learn and retain memories of their past and how this effects their behavior.

\subsection{Internal representations and anticipatory behavior}

There are two primary responses in which an organism may cope with a changing environment and the stresses that these induce: (i) by sensing environmental cues in order to anticipate and prepare for the change \cite{hilker2016priming}, or (ii) by activating multiple stress mitigating pathways in advance, a phenomenon that is called bet-hedging \cite{veening2008bistability}. 
\begin{marginnote}[]
\entry{Phenotypic Plasticity}{
Phenotypic plasticity refers to the capacity of genetically identical organisms
to exhibit different characteristics under varied environmental
conditions.
}
\end{marginnote}
Both of these types of responses rely on phenotypic plasticity, which is the ability of an organism to change its characteristics in response to environmental changes without altering its genetic content. Such responses come at a cost, as the cell has to reallocate limited resources in order to  synthesize the stress-mitigating metabolites  \cite{jun2018physiology_review}, and which may end up being of no use, if the anticipated changes did not happen. Biologist say that such strategies involve a fitness cost, since the price to pay for them is typically a reduction of growth. Thus in order to be viable, the fitness benefits of such strategies have to outweigh their cost, an issue that has been taken up in \cite{xueleibler2018phenotypic_plasticity}.
\begin{marginnote}[]
\entry{Fitness}{A generic attribute quantifying the adaptation of cells to an environment. For bacteria, rates of division or growth are fitness proxies. 
}
\end{marginnote}
Generally, anticipatory behavior requires the organism to have an underlying model that facilitates making predictions. Such models can be stochastic \cite{omer2025evolutionarylearning} or deterministic. For example, the posterior distribution in Bayesian inference serves the role of a model representing the environment, in this case the observed data. The relevance of representations via internal models  in biological systems has been discussed in \cite{tagkopoulos2008predictive, libchaber2020walking, eckmann2021dimensional, szathmary2022bayesian}.  

An example of an anticipatory strategy seen in bacteria is the switch from aerobic to anaerobic respiration after {\em E. coli} has been exposed to a change from ambient temperatures to $37^\circ$C. This metabolic switch is believed to be in anticipation of encountering an anaerobic environment, as would be the case when {\em E. coli} enters the oral cavity on its way to the gastrointestinal tract \cite{tagkopoulos2008predictive}. The environment of the gastrointestinal tract has many bacterial strains competing for limited resources, which thereby impedes colonization by new strains. Switching to an anaerobic metabolism early will convey an advantage to the invading strain relative to other invaders, by reducing the lag time that otherwise would ensue if the switching was made after the absence of oxygen had been detected. Moreover, the same study shows that such anticipatory responses can also be forgotten. By subjecting {\em E. coli} populations to environments where the change in temperature and oxygen levels were counter-correlated, i.e. both levels increased and decreased in parallel, resulted in selecting for strains that did well in these new conditions, but poorly in conditions mimicking the transition into the gastrointestinal tract. These findings demonstrate that the sequence of temperature change followed by metabolic switch is not hard-wired into the cells regulatory structure, but can be altered in principle. In the words of its authors the switching capability is learned as a result of ``evolution over geological time scales`` \cite{tagkopoulos2008predictive}. In the language of driven disordered systems, we may think of this is an emergent and reversible response produced by the training.

\subsection{The role of multi-stability, hysteresis and soft-modes in biological systems}

The Preisach model illustrates how a collection of bi-stable elements is capable of recognizing patterns of environmental changes by   mapping them into hysteron states. Such a system of bi-stable hysteretic switches can therefore filter environmental history. As we have seen, self-organization by environmental fluctuations can lead to the emergence of soft-spots, which, once they have become persistent, can recognize environmental histories, like the Preisach hysterons. 

Hysteron-like behavior is found in metabolic systems and comes in the form of molecular switches. One well-studied molecular switch with hysteresis is the  {\em lac} operon activation in {\em E. coli} \cite{ozbudak2004multistability} (see side bar). 
\begin{marginnote}[]
\entry{Operon}{
Collection of genes whose transcription is controlled by a common promoter region so that their proteins are synthesized together.}
\end{marginnote}
Sugars are the main energy source of bacteria. The {\em lac} operon is a set of genes that when transcribed permits the metabolisation of complex sugars such as lactose, when the preferred simple sugar glucose is not available. The activation of the {\em lac} operon comes at a cost, since some  resources have to be diverted into the synthesis of  proteins necessary to metabolize these complex sugars. Lambert and Kussell  performed experiments revealing HDB and memory formation when {\em E. coli} is repeatedly switched between glucose- and lactose-rich environments \cite{lambert2014memory}. It is thus conceivable that a metabolic system with many interacting hysteretic molecular switches would be capable to transit between metabolic states in a way that is highly sensitive to certain patterns of environmental fluctuations, giving thereby rise to complex HDB. 

\begin{textbox}[h!]\section{Bistability and hysteresis in gene regulation: the {\em lac} operon}
Metabolically, bistability (multistability) refers to genetically identical organisms capable of being in two (or more) distinct states, with the actual state depending on additional factors such as availability of nutrients. The switch between these states may be hysteretic with a positive feedback loop being often involved. The {\em lac} Operon in {\em E. coli} is a classic and well-studied example. The lac Operon controls the transcription of a set of genes that are necessary for the metabolization of complex sugars such as lactose. By default, transcription is repressed by a repressor protein encoded elsewhere on the genome. In particular lactose, when present, will cause deactivation of the repressor so that the genes of the lac operon can be transcribed and lactose metabolization is activated. Among the proteins synthesized, one of them enhances the deactivation of the repressor giving rise to a positive feedback loop.  The switching of the lac Operon is hysteretic, it becomes activated when lactose levels are sufficiently large, but will deactivate only if it has fallen below a lactose level much lower than what was needed for activation (for details see \cite{ozbudak2004multistability, lambert2014memory, phillips2020molecular}).   
\end{textbox}

\subsection{The role of dimensional reduction in biology and dose-response curves}

We have seen how the large configurational degrees of freedom of a system, when subjected to a one-parameter deformation under athermal quasistatic condition can be represented by a $t$-graph. The topology of the $t$-graph yields a minimal description from which properties like mechanical reversibility can be retrieved. This is an (extreme) example of dimensional reduction, where the quantities of interest are captured by a number of degrees of freedom that is much smaller than the available ones, meaning that  many  degrees of freedom are slaved to a few. This is the idea underlying a slow manifold and its soft-modes \cite{russo2025softmodes}. 

In the $t$-graph description, the underlying mesostates are not structureless. Referring to Fig.~\ref{fig:tgraph-intro}(a)  they represent a collection of microscopic configurations that transform continuously into each other as a deformation parameter $\varepsilon$ is varied. Macroscopic properties such as the external stress response to an applied strain evolve in a similar continuous manner within a mesostate, as the response is slaved to the particle configurations. These are the elastic branches, some of which were depicted in \ref{fig:amorphous-intro}(b).

Slow manifolds and their role in biology have been discussed extensively in a recent review  \cite{russo2025softmodes}. Along the same lines, the importance of dimensional reduction and continuous responses in biology has been revisited in  \cite{eckmann2021dimensional}. One of the manifestations of dimensional reduction in biological modeling are the ubiquitous {\em dose-response (DR) curves}. DR-curves describe the continuous dependence of quantities of interests on a set of control parameters. 
They play a role similar to the elastic branches of amorphous solids, which describe the stress response to strain deformation. 
DR-curves emerge for example in the description of metabolic switches \cite{ozbudak2004multistability, lambert2014memory}, but also in models of antibiotic resistance evolution \cite{roemhild2018cellular}, such as the tradeoff-induced landscape model \cite{Das2019}. The latter model exhibits hysteresis which can be linked to the Preisach model \cite{daskrugmungan2022}. Moreover, the complex adaptive response to a rapid change of antibiotic concentrations bears resemblance to stress relaxation in amorphous solids \cite{das2024biphasic}.

\subsection{Mechanical annealing as evolution in silico}

The self-organization of driven disordered systems under mechanical annealing mimics the evolutionary processes of mutation and selection in changing environments~\cite{Bak-Sneppen-PRL93, daskrugmungan2022, jaeger2024training}. In the context of annealing a sheared amorphous solid, we can think of the soft-spots as mutations, while the annealing by an externally applied shear -- cyclic or random --  emulates a selection process in which soft-spots are created and destroyed. The self-organization under annealing is a search for a collection of persistent soft-spots that can be repeatedly triggered in full or in subsets as the driving cycle or its subcycles are traversed, cf. Fig. \ref{fig:tgraph-intro}(d). As we have seen, this is the evolutionary process that leads to the emergence of mechanical reversibility and memory formation. Can  we regard mechanical annealing as being analogous to evolution, and if so, what insights can such an analogy provide?   

The evolution of organisms under mutation and selection is governed by a fitness landscape. Fitness is a proxy for the ability of genotypes to grow in an environment with evolution selecting for higher fitness and organisms evolving towards peaks of the landscape \cite{Wright1932, wrigth1982shifting, Weinreich2005,deVisserKrug2014}.  A candidate for fitness in the case of the sheared amorphous solid is the gain in elasticity as a result of reduction of energy dissipation, and hence diminishing of plastic activity under mechanical annealing \cite{ kumar2024self}. The self-organization of the amorphous solid under deformations, can then be regarded as that of a population of soft-spots that evolves by selection and mutation, according to  some appropriately defined fitness. 

The aforementioned work by Tagkopolous et al. \cite{tagkopoulos2008predictive} on the anticipatory metabolic switch in {\em E. coli} under temperature changes provides an example where this connection is explored in the other direction. The authors also simulate the evolution of biochemical networks whose properties changes according to empirically-motivated update rules and depending on the availability of fluctuating  environmental resources. They find that anticipatory behavior can be learned and subsequently forgotten under changes of environmental conditions. Analogies between the evolution of the biochemical networks and disordered solids under fluctuating environments are at hand. Mesoscale models of disordered solids, such as the QMEP model considered here, constitute attractive minimal models with a dynamic disorder landscape where such connections can be explored in  further detail.     

\section{OUTLOOK}
The following quote from Libchaber and Tlusty \cite{libchaber2020walking} describes well what may lie ahead of us:
{\em By construction, the memory cascade model embraces an “evolutionary ergodic principle” (or
evolutionary Occam’s razor): Every feasible mechanism that is not forbidden by hard physical
constraints will eventually be explored by evolution.} In this review, we have given an exposition of how physical systems with multi-stability and hysteresis respond to external stimuli and encode that history into their internal state. We propose here that such self-organization, coming "for free" as it were,  offers the possibility of facile pathways for evolvability and constitutes an appealing paradigm for understanding how biological organisms may sense and adapt to fluctuating environments. Pursuing analogies with physical systems capable of forming memories offers an exciting approach to comprehending the organization and evolution of simple enough biological systems.

\section*{DISCLOSURE STATEMENT}
The authors are not aware of any affiliations, memberships, funding, or financial holdings that
might be perceived as affecting the objectivity of this review. 

\section*{ACKNOWLEDGMENTS}
The authors would like to thank Taylan Cemgil, J\"urgen Horbach and Mukund Thattai for a critical reading of  this manuscript. They acknowledge the many useful insights gained from extensive discussions with their collaborators over the years. 
MM would like to thank Joachim Krug and Arjan de Visser for making him part of their experimental collaboration on antibiotic resistance evolution in {\em E. coli}. In particular, he acknowledges the many conversations with Dr. Rotem Gross, which provided a decisive stimulus for getting immersed into the amazing world of microbiology. 
MM's research was primarily supported by the Deutsche Forschungsgemeinschaft (DFG, German Research Foundation) under Projektnummer 398962893, and partly under  DFG project CRC 1310 {\em Predictability in Evolution}. 
SS's research is supported by SERB (ANRF) (India) through the JC Bose Fellowship (JBR/2020/000015) SERB, DST (India) and SUPRA project number SPR/2021/000382.

%

\bibliographystyle{dfg}%
\bibliography{AmorphNets_extnd_v2}%

\end{document}